\documentclass[twocolumn]{aa}
\usepackage{graphics}
\input{epsf}
\newcommand{\bfo}[1]{\mbox{\boldmath $#1$}}
\def\bvarphi{\mbox{\boldmath $\varphi$}}

\begin{document}
\newcommand{\beq}{\begin{equation}}
\newcommand{\eeq}{\end{equation}}
\def\la{\hbox{\raise.35ex\rlap{$<$}\lower.6ex\hbox{$\sim$}\ }}
\def\ga{\hbox{\raise.35ex\rlap{$>$}\lower.6ex\hbox{$\sim$}\ }}
\def\runit{\hat {\bf  r}}
\def\phunit{\hat {\bfo \bvarphi}}
\def\zunit{\hat {\bf z}}
\def\beq{\begin{equation}}
\def\eeq{\end{equation}}
\def\beqa{\begin{eqnarray}}
\def\eeqa{\end{eqnarray}}
\def\sub#1{_{_{#1}}}
\def\order#1{{\cal O}\left({#1}\right)}
\newcommand{\sfrac}[2]{\small \mbox{$\frac{#1}{#2}$}}
\title{Hydrodynamic stability of rotationally supported flows:
Linear and nonlinear 2D shearing box results
    \thanks{Research supported by
    the Israel Science Foundation, the Helen and Robert Asher
    Fund and the Technion Fund for the Promotion of Research} }

\author{O.M. Umurhan \and O. Regev}

\offprints{O.M. Umurhan, \email{mumurhan@physics.technion.ac.il}}

\institute{Department of Physics, Technion-Israel Institute of
Technology,
  32000 Haifa, Israel}

\date{Received April 1, 2004 / Accepted August 3, 2004}
\titlerunning{Stability of rotating circumstellar flows}

\abstract{ We present here both analytical and numerical results
of hydrodynamic stability investigations of rotationally supported
circumstellar flows using the shearing box formalism.  Asymptotic
scaling arguments justifying the shearing box approximation are
systematically derived, showing that there exist two limits which
we call small shearing box (SSB) and large shearing box (LSB). The
physical meaning of these two limits and their relationship to
model equations implemented by previous investigators are
discussed briefly. Two dimensional (2D) dynamics of the SSB are
explored and shown to contain transiently growing (TG) linear
modes, whose nature is first discussed within the context of
linear theory.  The fully nonlinear regime in 2D is investigated
numerically for very high Reynolds (Re) numbers. Solutions
exhibiting long-term dynamical activity are found and manifest episodic
but recurrent TG behavior and these are associated with the
formation and long-term survival of coherent vortices. The
life-time of this spatio-temporal complexity depends on the Re
number and the strength and nature of the initial disturbance.
The dynamical activity in finite Re solutions ultimately decays
with a characteristic time increasing with Re.  However, for large
enough Re and appropriate initial perturbation, a large number of TG
episodes recur before any viscous decay begins to clearly manifest itself.
In cases where Re = $\infty$ nominally (i.e. any dissipation resulting
only from numerical truncation errors),
the dynamical activity persists
for the entire duration of the simulation (hundreds of box
orbits).
Because the SSB
approximation used here is equivalent to a 2D incompressible flow,
the dynamics can not depend on the Coriolis force.
Therefore, three dimensional (3D) simulations are needed in order to decide if this
force indeed suppresses nonlinear hydrodynamical
instability in rotationally supported disks in the shearing box
approximation, and if recurrent TG behavior can still persist in three dimensions
as well - possibly giving rise to a subcritical transition to long-term
spatio-temporal complexity.

\keywords{accretion, accretion disks -- hydrodynamics --
instabilities -- stars: formation --
  stars: cataclysmic variables -- numerical methods}}
  \maketitle

\section{Introduction}

Accretion of matter, endowed with a significant amount of angular
momentum, onto relatively compact objects, is thought to occur in a
variety of important astrophysical systems, including close binary
stars, young stellar objects and active galactic nuclei. If the
accreting matter is able to cool efficiently enough, the resulting
configuration of the flow is a vertically thin, almost-Keplerian
(that is, essentially rotationally supported),
accretion disk. The accretion process (transport of mass inward)
can occur only if some dissipative mechanism is present, giving rise
to transport of angular momentum outward. Since the microscopic
viscosity of the fluids in question is very small, and thus the
Re numbers of such flows are literally astronomical,
anomalous transport of some kind must be invoked to account for
the time scales and magnitude of the energy output of the
accreting sources. The pioneering works of Shakura \& Sunyaev (1973)
and Lynden-Bell \& Pringle
(1972) suggested that the needed
anomalous transport can be naturally achieved if the accretion
flow is turbulent and, moreover, proposed to circumvent the
complexities and our lack of understanding of turbulence by
introducing the now famous $\alpha$ model of accretion disks
(in the language of turbulence modeling it is "a zero-equation,
one-parameter model"). This simple idea has been extremely fruitful,
giving rise to a large number of successful interpretations of
observational results (see e.g. Pringle 1981 and Frank, King \& Raine
2002 for reviews).

The theoretical question of the physical origin of accretion disk
turbulence has however remained essentially unanswered until
Balbus \& Hawley (1991) discovered that weak magnetic fields
destabilize differential Keplerian rotation, i.e.
the magneto-rotational instability (Velikhov 1959, Chandrasekhar 1960),
hereafter MRI, can operate in rotationally supported magnetized accretion disks,
rendering them linearly unstable. This instability was subsequently
recognized as the source of accretion disk magneto-hydrodynamic (MHD)
turbulence, giving rise to the needed angular momentum transport
and also to dynamo action (Hawley, Gammie \& Balbus 1995; Brandenburg
{\em et} al. 1995). For a recent review of this topic see Balbus (2003).
The discovery by Balbus \& Hawley is very important and significant
because it demonstrates the existence of linear instability
in accretion disk models with angular velocity profiles
that are otherwise stable according to the Rayleigh criterion.
Although some ideas have been put forward in this context, e.g.
the possibility of an instability resulting from the angular velocity being
nonconstant on cylinders (Knobloch \& Spruit \cite{knobloch},
Klu\'zniak \& Kita 1997, Regev \& Gitelman 2002,
Urpin 2003) or more general baroclinic instabilities (Klahr \&
Bodenheimer 2003), no hydrodynamic instability of any kind has ever been
{\em explicitly} shown to exist.
In the absence of a known linear hydrodynamic instability, the turbulent
viscosity needed for angular momentum transport in accretion disks
has acquired in the community a somewhat "mysterious" character,
similar to other ignorance-driven concepts in the history of
astrophysics (and physics). Thus the MRI has been generally accepted
with considerable enthusiasm and, as it often happens in cases where
a successful idea is advanced, raised to the level of an exclusive
paradigm. Moreover, the very possibility of the occurrence
of hydrodynamic turbulence (resulting from linear or even nonlinear
hydrodynamical instabilities) has been questioned by
Balbus, Hawley \& Stone (1996) (see also the detailed study
of Hawley, Balbus \& Winters 1999), hereafter BHSW, on the basis
of the results of fully nonlinear finite-difference numerical simulations
in the 3D shearing box approximation (see below).

The extreme displeasure felt by most astrophysicists
with the absence of linear hydrodynamical disk instabilities,
and their reluctance in accepting that disks may still be
hydrodynamically turbulent, seems to be
detached from what has been known for the past century on
hydrodynamical stability of shear flows.
It is well known (see e.g. Joseph 1972, Drazin \& Reid 1983)
that the linear stability properties of essentially all the canonical
shear flows, rotating or not, do not reproduce well their laboratory behavior.
In almost all such flows transition to turbulence is {\em subcritical},
that is, it occurs at Reynolds numbers (Re) that are significantly lower
(and also dependent on the perturbation strength) than the critical values
for linear instability ${\rm Re_c}$.
In some instances, like the pipe Poiseulle, plane Couette
or Taylor-Couette (in the narrow gap limit) flows, this problem
is particularly acute as linear analysis of perturbations on
these basic flow profiles predicts
stability (even decay) of infinitesimal perturbations, that is,
formally ${\rm Re_c}=\infty$. Laboratory experiments
demonstrate, however, full-on transition to turbulence via some
instability mechanism that is neither fully understood nor even
uniquely identified. Thus various ideas in this context
are generally referred to as transition {\em scenarios}
(see below).

On the theoretical side, it has been known since the work of
Orr (1907) that even though a basic shear flow may
be linearly stable, some perturbations (i.e.
initial conditions) may exhibit significant transient growth (TG)
within the linear regime, before ultimately decaying. This fact
prompted a number of researchers to examine the possibility of
{\em subcritical} transition to turbulence, with
the linearly stable flow finding a way to {\em bypass}
the usual (i.e. via linear instability) route to turbulence. We are
unable to mention here all the contributions on this subject
(in particular, those published before the early 1990s) and refer the interested reader to
one of the reviews cited below, but only note that an application
of TG to an astrophysical problem has already been implemented by Goldreich
\& Lynden-Bell (1965).

A significant step forward in the advancement of the bypass
transistion idea has been made in the early 1990s. It came with
the clarification that TG, and thus the possibility of a bypass
transition, is the result of the linearized system's operator
properties (Butler \& Farrell 1992, Reddy \& Henningson 1993). The
essence of this idea lies in the observation that linear stability
analysis of shear flows gives rise to a nonnormal operator(i.e.
an operator which does not commute with its adjoint),
that is, one whose eigenfunctions are not necessarily orthogonal.
This rather elementary mathematical fact may have profound
consequences for hydrodynamic stability, because it explains how
TG may naturally occur in linearly stable basic shear flows (see
e.g. Trefethen {\em et} al. 1993). This approach is frequently
referred to as {\em nonmodal} because the linear stability
analysis required to find TG cannot proceed via the usual modal
analysis, involving a boundary value problem and must consist of
the examination of the original initial value problem.  If
this TG is strong enough then nonlinear interactions may come into
play.   This then could further lead to some kind of
"nonlinear instability" creating
a sustained complex dynamical state in which TG events
persistently recur and resulting in a situation resembling turbulence.
We shall refer to this phenomenon as {\em recurrent transient growth} (RTG).
The recent review by Grossmann (2000) contains an excellent
account of these ideas. We refer the reader also to the
comprehensive modern book on shear flow instabilities
(Schmid \& Henningson 2001) and relevant references within,
where a detailed account (a full chapter) on the different
ideas on transition to turbulence in shear flows are detailed.

The above developments recently have been  brought to the attention of
the astrophysical community  by Ioannaou \& Kakouris (2001) and
Chagelishvili {\em et} al. (2003) in the context of accretion disks.
These two studies are quite different in their approach
- the former examines global behavior of disturbances in an
accretion disk and the latter is a local analysis in the framework
of the {\it shearing box} (SB), also sometimes called the {\it shearing
sheet}, approximation - but they both utilize the basic idea
mentioned above. It appears that although these works have not
explicitly demonstrated that a turbulent state of an accretion disk
actually arises by purely hydrodynamical processes, they have convincingly
suggested that such a process is feasible. The purpose of this
(and subsequent) work is to carry this idea further and examine
in detail the nonlinear regime in various limits.

There is little doubt that whenever MRI occurs in a disk it leads to a
turbulent flow with enhanced angular momentum transport {\em outward}.
However, BHSW (see also Balbus 2003) put forth the
claim that no purely hydrodynamical process can destabilize a Keplerian
accretion disk, at least within the SB approximation. This is based on
fully {\em nonlinear} simulations (performed by BHSW) in which no (nonlinear) hydrodynamic
instability was shown to manifest itself as long as the angular velocity profile was
{\em linearly} stable according to the Rayleigh criterion.
Furthermore, the same simulations of BHSW also show
that the total kinetic energy of plane Couette flow
and rotating flows which are unstable according to
the Rayleigh criterion, grow essentially monotonically (after a
short transient).
By contrast, in Rayleigh-stable rotation profiles the kinetic
energy was observed to decay over the duration
of the runs (typically 5-10 box orbit times).\par
Since in the SB approximation it is the Coriolis term that actually
distinguishes between simple plane Couette flow and
the flow within a small
segment of a rotating disk (see below),
it is only natural that this force has been identified by BHSW
as the stabilizing agent.
\par
These numerical results certainly constitute a serious challenge
to the possibility of hydrodynamic turbulence in rotationally
supported disks. On the other hand, we can state at least five
observations which suggest that this matter is far from being settled.
\begin{enumerate}
\item In some astrophysical systems containing such disks the
conditions are such that MRI driven MHD turbulence is probably
impossible. For example, accretion disks in dwarf novae during
quiescence (Menou 2000) and around forming stars (Blaes \& Balbus
1994, Sano {\em et} al. 2000) seem to be too resistive to support
MHD turbulence. \item Recent laboratory Couette-Taylor experiments
in the narrow gap limit, with linearly stable rotational angular
velocity profiles (like in Keplerian disks) indicate the
development of turbulence in such flows (Richard 2001, see also
Richard  \& Zahn 1999). \item In a recent thoughtful examination
of the problem, Longaretti (2002) proposes that the above
conclusion reached by BHSW, about the role of the Coriolis force
in suppressing the subcritical transition (observed in
non-rotating plane Couette flows in the lab and in BHSW
simulations), may be premature. He speculates that their findings
may stem from the inability (of the simulations) to resolve some
of the essential dynamics within these rotating flows (see however
Balbus 2003). \item There exist numerical simulations of plane
Couette flows, which have revealed the details of what can be
called "subcritical turbulence" in these flows (Schmiegel \&
Eckhardt, 1997; Eckhardt {\em et} al. 1998). These calculations,
which are the only ones (as far as we know) of this kind and
extend for reasonably long integration times, indicate that above
a critical Reynolds number there appears dynamical activity,
resulting probably from repeated TG events, brought about by
nonlinear interaction and persisting much longer than the viscous
decay time, i.e. RTG. This nature of the subcritical bypass
transition appears to be very different from a "usual"
supercritical transition to turbulence via linear instability. In
particular the outcome critically depends on the initial
perturbation, in addition to the Reynolds number. Similar
dependence is observed in experiments (see Grossman 2000). It is
therefore quite surprising that in BHSW the non-rotating Couette
flow (which is linearly stable and hence must have a subcritical
transition and thus probably RTG) and the rotating Rayleigh
(linearly) unstable flows exhibit a similar perturbation energy
growth, at least during the relatively short duration of the BHSW
simulations and for a particular choice of the initial
perturbation. \item In relation to the last point, the simulations
of BHSW only consider one type of initial conditions to seed their
flows and they follow its evolution for, at most, a dozen orbit
times.  Perturbation spectra are surely myriad and it is possible
that conditions leading to a subcritical transition into
turbulence may have been overlooked by BHSW.

\end{enumerate}
\par
Thus, it seems that continued study of purely hydrodynamical
processes in disks still remains viable and worthwhile (in the words
of Balbus, Hawley \& Stone 1996) and, in particular, the issue
of hydrodynamical stability warrants perhaps a new look.

In this paper we will report on our first effort in this
direction. Our goal will be to critically evaluate, using high
resolution and long duration (hundreds of box orbit times, that
is, Kepler rotational periods at the radius of the box location)
numerical experiments in the nonlinear regime, whether the linear
TG mechanism is sufficient to induce
a bypass transition into a persistently dynamically active
state, like the one described in item (4) above. We shall employ spectral
methods (wherever possible) for our simulations, as these methods
are generally regarded in the fluid-dynamical community as being
more accurate and reliable than finite difference schemes. For the
time being we shall focus just on the nature of the instability
and the possibility of a sustained spatio-temporal complexity. The
issue whether in such a state there is an enhanced angular
momentum transport {\em outward} (as it must be in accretion
disks) is an important one (see Cabot 1966, Balbus 2003), but its
exploration must await a conclusive result on the instability
itself and thus we shall not deal with it in the present work.

This paper is organized in the following way. In Section 2 we
examine and discuss the nondimensional SB equations, as
appropriate for the case studies here, in which the size of the
box is much smaller than the disk's vertical scale height.
This is one of the limits of the
SB approximation, whose systematic and general derivation
is detailed in Appendix A. Section 3
deals with linear theory and in particular we apply there the
concept of sheared coordinates, with whose help the
results of Chagelishvili {\em et} al. (2003) (the TG of linear modes)
can be elegantly recovered. In section 4 we report the
results of nonlinear 2D numerical simulation in this limit of the
SB approximation. Although it has been recognized that
nonmodal TG may give rise to very large amplification in
3D, also 2D perturbations can grow quite significantly
(Trefethen {\em et} al. 1993).
In two-dimensional viscous incompressible flows any dynamical activity must
ultimately decay (see section 4.2 below), however for large enough Re numbers
we expect the decay times to be very long (since they should behave as $\propto
{\rm Re}^{2/3}$, Yecko  2004) so as to allow for the TG episodes to recur many times.

Thus, in this initial study we limit ourselves to
2D and we postpone to subsequent investigations (which are
already in progress) 3D simulations of this and the other
limit of the SB approximation.

\section{The Shearing Box Approximation Equations:}

Numerical simulations of rotationally supported flows at very large
Reynolds numbers must focus on only very small disk sections
if high spatial resolution is imperative. This is clearly the case when one
looks for possible instabilities and therefore most such studies
have been done within the shearing box (or sheet) approximation.
The essence of this approximation is local in approach, that is,
the equations are valid in a small
region (a Cartesian box) about a typical point in the disk.
In such a box, a steady flow consisting of a linear shear velocity
profile solves the equations and one can consider it as a basic flow
and perturb around it. The equations for the perturbations (linearized
or not) are then referred to as the shearing box (SB) approximation equations.
Most numerical simulations of Balbus, Hawley and collaborators in the MRI
context have been done within this formalism as have the recent studies, mentioned
above, on the feasibility of a hydrodynamical instability. Although
the SB approximation is not new (Goldreich \& Lynden Bell 1965) we
have not found in the literature a systematic derivation of this
approximation. Thus, in the purpose of resolving the confusion
(of the present authors and possibly also others) we give here in Appendix A
such a derivation, giving rise to two limits of this approximation.

In the first one, a box whose size is much smaller than the
smallest scale height in the disk (usually the one in the
vertical direction) is considered, and thus
the unperturbed state of linear shear may be considered homogeneous.
The perturbations are then incompressible and acoustics are ruled out.
Most previously relevant work have been done in
this approximation, which we call {\em small shearing box} (SSB).
For the sake of completeness we repeat here the non-dimensionalized SSB equations
derived in Appendix A.  In writing the equations below, we have deviated slightly
from the notation presented in Appendix A,
\beqa
 \nabla \cdot {\bf u} &=& 0,
\label{enssbox}
\\
\partial_t u + 2 A x \partial_y u - 2 \Omega_0 v +{\bf u} \cdot \nabla
u &=&
 -\frac{\partial_x p}{1+\rho'},
 \label{mxssbox}
\\
\partial_t v + 2 A x \partial_y v + 2 (\Omega_0+A) u +{\bf u} \cdot \nabla
v &=& - \frac{\partial_y p}{{1+\rho'}} , \label{myssbox}
\\
\partial_t w + 2 A x \partial_y w  +{\bf u} \cdot \nabla w &=&
- \frac{\partial_z p + z\rho'}{1+\rho'} , \label{mzssbox}
\\
\partial_t \rho' + 2 A x \partial_y \rho' +  {\bf u}\cdot \nabla \rho' &=& 0,
\label{msssbox}
\eeqa
and the identifications to the terms in Appendix A are: ${\bf u} \equiv {\bf u'}$,
 $u \equiv u'_x, v \equiv u'_u, w \equiv u'_z$.
We also specifically assume that the disk is exactly Keplerian which
means that $\Omega_0 = 1$ and $2A = -3/2$.  It also means, by (\ref{px}), that
$\partial_z P_b = -z$ and $\partial_x P_b = 0$.  Additionally, even though $\Omega_0 = 1$,
we retain this symbol
throughout the calculation for the purpose of flagging the Coriolis terms.
\par
In this paper we exclusively use the SSB approximation
and assume that the initial density disturbance is everywhere zero, i.e. $\rho'(t=0) = 0$.
From (\ref{msssbox}) it follows that
$\rho' = 0$ for all subsequent times.  Thus,  (\ref{msssbox}) becomes irrelevant
and from here on out it shall not be referenced.\par
The large shearing
box (whose size is of the order the disk thickness) equations are also given
in Appendix A and they will be treated in our future works.

\section{SSB linear theory - a review}


For the sake of economy
we reintroduce the parameter $q$ (with $\Omega_0 = 1$) as defined by (\ref{qdef})
to replace $- 2 A $ in the linearized expressions of (\ref{enssbox}-\ref{mzssbox}):
\beqa
\partial_x u + \partial_y v + \partial_z w &=& 0,
\label{ssblinener}
\\
(\partial_t - q x \partial_y)u - 2 \Omega_0 v &=&
-\partial_x p, \label{ssblinu}
\\
(\partial_t - q x \partial_y)v + (2 \Omega_0 - q) u &=&
-\partial_y p, \label{ssblinv}
\\
(\partial_t - q x \partial_y)w &=& -\partial_z p.
\label{ssblinw}
\eeqa
Since this flow is exactly Keplerian, $q = 3/2$.
As a point of reference: for solid body rotation $q = 0$, for
flows with constant specific angular momentum $q = 2$. These
linear equations describe simple incompressible flow with a linear
shear profile in a rotating frame: that is, linearized
rotating plane Couette flow (Nagata, \cite{nagata}).
\par

For {\em two-dimensional} disturbances
(i.e. when $\partial_z = 0$, $w = 0$), the equations
governing the linear evolution simplify.  With the vertical
component of vorticity defined by,
\beq \xi \equiv
{\partial_x v} - \partial_y u,
\label{2Dvorticitydef}
\eeq
it follows from (\ref{ssblinu}) and (\ref{ssblinv})
that it is conserved along the shear flow,
\beq
(\partial_t - q x \partial_y) \xi = 0.
\label{2Dvorticity}
 \eeq
Because of the incompressibility of the
2D flow, the velocity components may be derived from a stream
function, $\psi$, defined by
\beq
u = -\partial_y \psi, \quad v = \partial_x \psi,
 \label{Psirelations}
\eeq
and related to the vorticity via
\beq
\xi = \partial_x^2\psi + \partial_y^2\psi.
\label{xiPsi}
\eeq
We note also that the equations governing 2D
dynamics are insensitive to the background rotation state even though
the kinematics are sensitive to it. If the boundary conditions
are independent of rotation too (as is the case here), the
velocity distribution (that is the dynamics) cannot depend
on the Coriolis term.
This is a well known property of the governing equations and it is entirely
due to the incompressibility of the flow
which facilitates the streamfunction-vorticity formulation
(see e.g. Batchelor 1967, p. 178).
The resulting equation set, even
the nonlinear one, becomes mathematically equivalent to the set appropriate to
plane Couette flow. The results of all our 2D calculations are
based on solving equations (\ref{2Dvorticity}) and (\ref{xiPsi})
with periodic boundary conditions (in the sheared frame, see below)
and thus they do not depend in any way on the
Coriolis force.
\par
To avoid any confusion with language,
in this work we interchangeably refer to points with constant values of $x$
and varying values of $y$
as being in the {\em streamwise} direction
and we
refer to points with constant values of $y$
and varying values of $x$
as being in the {\em shearwise} direction.
\par
Note that for {\em three-dimensional} disturbances one can
obtain a single decoupled
equation for the radial velocity component $u$ after
manipulation of (\ref{ssblinener}-\ref{ssblinw}),
\beq
 (\partial_t - q x \partial_y)^2 (\partial_x^2 +
\partial_y^2 + \partial_z^2)u+ 2\Omega_0(2\Omega_0 - q)\partial_z^2 u = 0.
\label{eqnforux}
\eeq
Finally we point out that the same manipulations show that
the pressure is related to the radial velocity by
\beq
\partial_y^2 p + \partial_z^2 p = (\partial_t - qx \partial_y )\partial_x u -
(2\Omega_0-q)\partial_y  u.
\label{pressureuxrelationship}
\eeq
In contrast
to 2D dynamics, 3D dynamics feel the effect of rotation
through the explicit presence of the Coriolis term
$2\Omega_0(2\Omega_0-q)$ in (\ref{eqnforux}).
\par

We continue the analysis by reformulating the linear problem in
terms of {\it shearing coordinates} ( SC for short) in the same way
as implemented by Goldreich \& Lynden-Bell (1965).
To make sure our terminology is clear, we shall refer
to the usual (untransformed) coordinate formulation
of this problem to be in the {\it non-shearing coordinates} (NSC for short).
The NSC is what an observer would see and, as such, we interchangeably refer to it
as the {\it observer frame}.
\subsection{Shearing coordinates}\label{shearing_formalism}
The coordinates are written into a frame which is shearing exactly
as the background flow itself.  This new coordinate
system is formally defined as,
\beq X = x,\quad Y = y + q (t-t_{_0}) x
,\quad Z = z,\quad T = t-t_{_0}, \label{shearedmetrictransform}
\eeq
where $t_{_0}$ is some arbitrary reference time.
Derivative operators are replaced in the following sense,
\beq
\partial_t = \partial_{_T} + q
X {\partial_{_Y}}, \ \partial_x
= \partial_{_X} + q T \partial_{_Y}
, \ \partial_y = \partial_{_Y} ,
\ \partial_z = \partial_{_Z}.
\label{operatortransform}
\eeq
The 3D equations that result from
this transformation applied to equations (\ref{ssblinener}-\ref{ssblinw})
are,

\begin{mathletters}
\beqa
\left({\partial_{_X}} + q T{\partial_{_Y}}\right) u +
{\partial_{_Y}}v + {\partial_{_Z}}w  &=& 0,
\label{shearedcontinuity}
\\
\partial_{_T} u - 2 \Omega_0
v &=& -\left({\partial_{_X}} + q T
\partial_{_Y}\right) p,
\label{shearedmx}
\\
\partial_{_T} v + (2\Omega_0-q)u &=&
-{\partial_{_Y}}p,
\label{shearedmy}
\\
\partial_{_T} w &=& -\partial_{_Z} p.
\label{shearedmz}
\eeqa
\end{mathletters}
\par
The advantage in going over into
this reference frame is that we explicitly remove the shear
expression (which is proportional to $x$) from the governing
equations.  In return we receive a term that is proportional to time,
$T$, in the resulting initial value problem (IVP). Rewriting
(\ref{eqnforux}) and (\ref{pressureuxrelationship}) in the SC gives,
\beqa
 & & \partial_{_T}^2 \left[\left(\partial_{_X} + q T \partial_{_Y}\right)^2 + \partial^2_Y
+ \partial_{_Z}^2 \right] u  \nonumber \\
& & \qquad\qquad\qquad\qquad + \ 2\Omega_0(2\Omega_0 - q)
\partial_{_Z}^2  u = 0,
\label{shearedeqnforux}
\eeqa
and
\beq
\left(\partial_{_Y}^2  + \partial_{_Z}^2\right) p =
\left[\partial_{_T}(\partial_{_X} + q T \partial_{_Y})u - (2\Omega_0 - q) \right ]u.
\label{shearedpressuremxrelationship}
\eeq

We remind the reader that this coordinate
transformation is volume preserving because its Jacobian is exactly one.
\par
We investigate the linear behavior in terms of the Fourier components
of the disturbances.  This is a natural choice if we are to investigate
the case of periodic boundary conditions in SC
(a rather natural choice).  Thus, any relevant physical quantity, $f$,
has the form,
\beq
f(X,Y,Z,T) = f_{_{k\ell m}}(T)e^{ikX + i\ell Y + imZ} +  c.c.
\label{solutionsheared}
\eeq
Though disturbances are written here as 3D (for further studies)
we shall focus only on 2D behavior in this work. Consequently,
we may consider the dynamics within the
vorticity-streamfunction formulation which greatly simplifies the analysis.
Also, the $m$ subscript on all Fourier components will be suppressed hereafter.
\par
We now translate relationships (\ref{2Dvorticitydef}) - (\ref{Psirelations})
into the language of the SC to find,
\beqa
u &=& -\partial_{_Y} \psi, \quad
v =  \left(\partial_{_X} + q T \partial_{_Y}\right)\psi,
\label{sheared2Durelationships}
\\
\xi &=& \partial_{_X}^2 \psi + \left(\partial_{_X}  + q T \partial_{_Y}\right)^2\psi,
\label{sheared2Dxipsirelationships}
\eeqa
and
\beq \partial_{_T} \xi = 0. \label{sheared2Dvorticityevolution}
\eeq
By (\ref{sheared2Dvorticityevolution}) we immediately see that
$\xi_{_{k\ell}}(T) =
\xi_{_{k\ell}}(T=0) \equiv \tilde\xi_{_{k\ell}}$ is a time invariant quantity.
It is, therefore, straightforward to demonstrate that
introduction of the solution in the form (\ref{solutionsheared})
into (\ref{sheared2Durelationships}-\ref{sheared2Dxipsirelationships})
gives,
\beqa
\psi_{_{k\ell}}(T) &=& -\tilde\xi_{_{k\ell}}\frac{1}{(k + q T \ell)^2 + \ell^2},
\\
 u_{_{k\ell}}(T) &=&  -\tilde\xi_{_{k\ell}}\frac{i\ell}{(k + q T \ell)^2 + \ell^2},
\\
 v_{_{k\ell}}(T) &=&
\tilde\xi_{_{k\ell}}\frac{i(k + q T \ell)}{(k + q T
\ell)^2 + \ell^2}
.
\eeqa
The above solutions are the same as those derived by
Chagelishvili {\it et} al. (\cite{chag}) and they contain
the possibility of TG for some initial conditions.
Specifically, a maximum in the amplitude of $
u_{_{k\ell}}$ exists when $T = T_{\rm{max}}= -k/(q\ell)$.  This maximum is
achieved
for $T>0$ only for combinations of $k$ and $\ell$ in which $k\ell
< 0$ and we will refer to these modes as {\em transiently growing}
or {\em leading}. On the other
hand, those modes in which $k\ell >0$ always show decay for $T>0$ and
we refer to these as {\em trailing}.\par
Similarly, a maximum in $v_{_{k\ell}}$ occurs at $T = T_{\rm{max}}
+1/q$. Nevertheless all linear solutions asymptotically
die away as $T\rightarrow\infty$,  where in
particular,
\beq u_{_{_{k\ell}}} \sim \frac{1}{T^2},\quad v_{_{k\ell}} \sim
\frac{1}{T}, \qquad {\rm as}, \ T\rightarrow \infty. \eeq
\par
\subsection{Energy}

We define the domain integrated disturbance kinetic energy
\footnote{Since the kinetic energy is the only form of energy in our
problem, the "kinetic" qualifier is suppressed
hereafter.}
per Fourier mode to be
\beqa
E_{_{k\ell}} &=&  \frac{1}{2}\int\Bigl\{ \left(u_{_{k\ell}}e^{i(kX
+ \ell Y)} + c.c.\right)^2 \nonumber \\
&+& \left(v_{_{k\ell}}e^{i(kX + \ell Y)} + c.c.\right)^2 \Bigr
\}dX dY  , \eeqa which, after using the derived solutions above,
reduces to \beq E_{_{k\ell}} = \frac{|\tilde\xi_{_{k\ell}}|^2}{(k
+ q T\ell)^2 + \ell^2} L_xL_y,
\label{linear2Denergy}
\eeq
where $L_x,L_y$ are the periodic scales in the respective $X$ and $Y$
directions. \par The transient growth of energy is demonstrated
very clearly in the expression for $E_{_{k\ell}}$: like for the
$u_{_{k\ell}}$ disturbances - a mode will achieve a maximal growth
of its energy at $T = T_{\rm{max}}$ and this happens for $T>0$
only if $k/\ell < 0$.\par
\begin{figure}
\begin{center}
\leavevmode \epsfysize=5.cm
\epsfbox{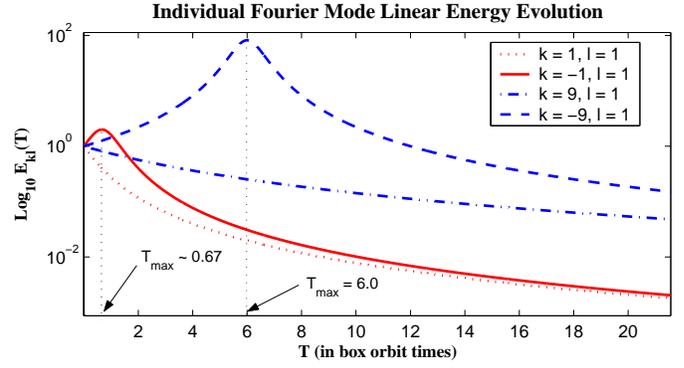}
\end{center}
\caption{{\small  The behavior of the linear mode energy, $E_{_{k\ell}}$,
with respect to time (\ref{linear2Denergy}) in the SC frame is
plotted for several $k$ and $\ell$ Fourier modes.  Each mode
begins with the same energy at $T=0$.  Transient growth
is strongest when the value of $|k/\ell|$ is largest (see the
definition of $\vartheta_{k\ell}$ in Eq. \ref{amplificationfactor}).
The times of maximal growth ($T = T_{{\rm max}}$) in
the TG modes are also labelled.}}
\label{fig:linear_tg_growth_theory}
\end{figure}

 Let $\vartheta_{_{k\ell}}$ measure the
amplification of the energy of a Fourier mode at times $T=0$ and
$T=T_{\rm {max}}$, \beq \vartheta_{_{k\ell}} =
\frac{E_{_{k\ell}}(T_{\rm {max}})}{E_{_{k\ell}}(0)} =
\frac{k^2}{\ell^2} + 1. \label{amplificationfactor} \eeq This
definition most clearly shows that the greatest amplification
occurs for single modes whose ratio $k/\ell$ is greatest. Typical
behavior of this transient growth is displayed, for several values
of $k$ and $\ell$, in Fig. \ref{fig:linear_tg_growth_theory}.
\par
Letting $E_d$ designate
the domain integrated energy of the disturbances,
it is a simple matter to write it as the sum
of the individual mode disturbance energies $E_{_{k\ell}}$,
\beq E_d =
\sfrac{1}{2}\int{\left( u^2 + v^2 \right )
dX dY} = \sum_{k,\ell} E_{_{k\ell}}.
\label{total2Ddenergy}
\eeq
Note that this general
expression of the domain integrated disturbance energy is valid in the
nonlinear case - the only difference is that
the (constant) mode amplitude, $\tilde\xi_{_{k\ell}}$,
appearing in (\ref{linear2Denergy}) will be
replaced by the nonlinearly time varying amplitude
$\xi_{_{k\ell}}(T)$ instead (see below).
The domain integrated disturbance energy will follow the same prescription as in
(\ref{total2Ddenergy}).

\section{2D Nonlinear Behavior}

\subsection{Equations Solved and Numerical Method}
We consider fully nonlinear two dimensional disturbances
(\ref{enssbox}-\ref{myssbox} with
$\partial_z = 0, w = 0$ and $\rho' =0$) of the
fluid with periodic boundary conditions in the SC frame.
The equations of motion for the disturbances are evolved using the
streamfunction-vorticity formulation discussed in the previous sections.
The full nonlinear equations in the NSC frame are,
\beqa
\partial_t \xi - q x \partial_y \xi + \partial_y\psi\partial_x\xi
&-& \partial_x\psi\partial_y\xi = \nonumber \\
& & {\rm Re}^{-1}\left (\partial_x^2\xi + \partial_y^2\xi\right),
\label{fullxieqnnsc}
\\
\partial_x^2 \psi + \partial_y^2\psi
&=&
 \xi.
 \label{streamvortyrelnsc}
\eeqa
When the transformations defined by
(\ref{shearedmetrictransform}) are applied, the equations in the SC frame are,
\beqa
\partial_{_T} \xi + \partial_{_Y}\psi ( \partial_{_X}\xi &+& q T\partial_{_Y}\xi)
-\partial_{_Y}\xi\left(\partial_{_X}\psi + q T\partial_{_Y}\psi\right)
= \nonumber \\
&& {\rm Re}^{-1}\left[\partial_{_Y}^2\xi + \left(\partial_{_X} +
q T\partial_{_Y}\right)^2\xi\right],
\label{fullxieqn}
 \\
\partial_{_Y}^2\psi + (\partial_{_X} &+& q T\partial_{_Y})^2\psi
= \xi,
 \label{streamvortyrel}
\eeqa
where we have added here, for the first time, an explicit viscous term
to the equations. Specifically, the equations contain now a
parameter (the Reynolds number). The purpose of this inclusion is
to investigate how the flow behavior depends on Re and what influence
Re has on the transition to the RTG state.  For the most part, however,
we report on inviscid results here, that is, with formally
Re $= \infty$ in the above equations.  We note that the numerical procedure
implemented (see below) has a certain amount of artificial dissipation associated with it
so that though the Re number may be infinite, it is so only nominally.
\par
Doubly periodic boundary conditions are imposed upon all the quantities in the SC frame.
This is identical with the conditions used by Hawley, Gammie \& Balbus (\cite{hgb}).
\par
Equations (\ref{fullxieqn}-\ref{streamvortyrel}) are solved using
standard Fourier spectral methods (Canuto,  et al. \cite{canuto})
except for a number of details discussed below.  Runs were
performed usually on specific periodic domains ($L_x$,$L_y$) =
($\pi,2\pi$). Because the nonlinearities are quadratic, a 3/2
dealiasing rule is imposed (i) at each time step, after the nonlinear terms
are computed and (ii) prior to each remapping event (see below).
Aliasing problems can be especially hazardous here
because one can
inadvertedly transfer power from decaying (trailing) modes into
transiently growing (leading) modes and create a situation in which
there is spuriously generated RTG behavior.  The dealiasing procedure
outlined above satisfactorilly avoids this problem.\par
The Fourier resolution typically
utilized 256 modes in $Y$ and $512$ modes in $X$.
Convergence of the numerical scheme was
verified by comparing the results of representative runs with
those of an even higher resolution simulations ({\it e.g.} at
$512\times 1024$).
These numerical tests convince us that the 256$\times$512 spatial
resolution is satisfactory for our purposes at this stage.

\par
The governing PDE's were temporally evolved using a modified
Crank-Nicholson method described in Appendix
{\ref{numerical_methods}}.

The domain in the $x=X$ direction nominally lies between $X=0$ and
$X = L_x$.  In the NSC frame, the background flow is not moving at
$x=0$ while it moves with speed $q L_x$ at $x = L_x$. Consequently
this means that the shear carries all the $y$ points at $x=L_x$
back to their initial starting position after a time $T_{rm}$
defined to be \beq T_{rm} = \left|\frac{L_y}{q L_x}\right|.\eeq
However, as noted in Hawley, Gammie \& Balbus (\cite{hgb}),
points, as carried by the shear, are not always periodic in $x$:
they are so only after every integer multiple of $T_{rm}$.
\par Thus, following the general prescription provided for by Cabot
(\cite{cabot}), the computational domain is {\em remapped} after
every $T_{rm}$ period of time according to the following prescription:
\begin{enumerate}
\item At $T=t=0$ initial vorticity (or streamfunction) data
are assumed to be known and given {\it in the NSC frame}.
At this time the NSC ($x,y,t$) and SC ($X,Y,T$) are
coincident.
\item Given this coincidence of both coordinate systems, the
initial NSC data are used as the seed vorticity
(or streamfunction)
that evolve by (\ref{fullxieqn}) and (\ref{streamvortyrel}),
the equations in the SC, up to time $T=T_{rm}$.

\item Because only at $T=T_{rm}$ (and every integer
multiple thereof ) are solutions exactly periodic
in the $x$ direction of the NSC frame, at $T=T_{rm}$
the solution in the SC frame is mapped back into the
NSC frame by applying the inverse transformation of
(\ref{shearedmetrictransform}) onto the modes in Fourier space
according to the procedure in Appendix C, section 2.
This act is what we will refer to as {\it remapping}.
\item We can treat the
mapped, exactly periodic, solution in the NSC frame as the
new initial condition to evolve
(\ref{fullxieqn}-\ref{streamvortyrel}) which, of course,
is in the SC frame.  Time in SC is reset
to zero, {i.e.} $T=0$.  The procedure repeats with item 2.
Note that though $T$ is reset, time in the NSC still continues
onward.  For instance, at the $n^{{\rm th}}$ remapping, $T=0$ while
$t=nT_{rm}$.
\end{enumerate}

This procedure simply amounts to taking a solution that has been evolved
in the SC and re-expressing it in terms of the coordinates of a
normal (that is, in NSC) observer.  Then the data as seen by the normal
observer is used again as initial conditions for the next spate of
evolution in the SC frame.
It should be noted that if there were infinite
resolution at our disposal, the remapping procedure mathematically effects nothing since
it is a simple coordinate transformation.
The rationale for applying the remapping procedure
is technical: it involves issues concerning finite resolution,
periodic boundary conditions, and the ability of resolving coherent structures
for an extended period of time in the SC frame.
For more details we refer the reader to Cabot (\cite{cabot}) and to the discussion presented
in Appendix C.
Prior to
each reemapping act, the state of the solution is
padded using the similar 3/2 dealiasing procedure (see above).
Failing to do this can introduce
an aliasing error in which a decaying (trailing) Fourier
mode can spuriously turn
into a transiently growing (leading) Fourier mode
({\em cf.} end of section 3.1).  Nonetheless, we have also performed
a comparison run to see whether or not the recurrent transient growth
we observe in the nonlinear solutions (see section 4.3) indeed persists
when remapping is not done.  Aside from slight differences in the evolution
of the total disturbance energy and the differences in the enstrophy
(when remapping is applied enstrophy decay
is expected and observed while when no remapping is performed
enstrophy conservation is predicted and observed),
the observed recurrent dynamical
phenomena persists in the non-remapped case just as it does in the
simulations where remapping is done.
\par
We define the parameter $\varepsilon$ to be a rough
measure of what we shall call the {\it turbulence intensity} of the system.
In particular we identify it to be the ratio of
the domain integrated energy contained in fluid disturbances
({\it i.e.} the energy $E_d$ as defined by Eq. \ref{total2Ddenergy})
to the domain integrated energy contained in the steady shear flow
itself.  The latter is the domain integral defined by
\beq
E_{shear} = \frac{1}{2}\int {\bf U}^2 dx dy = \frac{1}{2}\int q^2x^2 dx dy =
\frac{1}{6}q^2 L_x^3 L_y.
\eeq
The turbulent intensity is then formally written as,
\beq
\varepsilon =
\frac{E_d}{E_{shear}} = \frac{6 E_d}{q^2 L_x^3 L_y}
\eeq
\par
All simulations in this work, unless otherwise noted, begin with white noise
in the vorticity field with a value of $\varepsilon$ always less than $0.01$.
That is to say, the energy in the disturbances are always less than
one percent of the total energy in the steady shear.
\par
\subsection{Disturbance energy and enstrophy evolution}
To gain a certain amount of intuition for the underlying processes and to monitor
the performance of our numerical technique we consider the evolution
of the domain integrated energy of the disturbances.  It is a straightforward matter
to show that in the SC the following holds,
\beq \frac{d E_d}{dT} =
\left .q\int  u v dXdY
- Re^{-1}\int |\tilde\nabla u|^2 dXdY
\right. ,  \label{reynolds-orr}
\eeq
where the operator $\tilde\nabla$ is given as $\{\partial_X + qT\partial_Y\}{\bf \hat X}
+ \partial_Y{\bf \hat Y}$ and the integrals are taken over the periodic domain.
The  result in (\ref{reynolds-orr}) is known in more general terms as the Reynolds-Orr relationship.
When Re is finite it clearly demonstrates the decaying behavior due to viscosity
since the second term on the RHS of (\ref{reynolds-orr}) is a negative definite quantity.
In the limit where Re is so large that the viscous decay time scale is very long, the
evolution of $E_d$ is governed by the first term on the RHS of (\ref{reynolds-orr}).
It says that for $q>0$ ($<0$) there will be a rise in the integrated kinetic
energy of the domain if there exists a positive (negative) correlation between
the shearwise and streamwise velocities. From (\ref{reynolds-orr}) it can be shown
that the instantaneous
relative growth rate of the disturbance energy, that is,
$d \ln E_d /dt$ is independent of the amplitude (see Hennigson 1996). Thus the growth
rate of a {\em finite} amplitude disturbance is essentially given by mechanisms present
in the the linearized equations.
Without linear growth mechanisms there can be no
growth even in the nonlinear regime. Nonlinear processes can only shift power
among the modes and this, as we shall explain below, is a key feature in understanding
our results.
\par
Additionally we also consider the {\it enstrophy},
\beq
{\cal Z} \equiv \int \sfrac{1}{2}\xi^2 dXdY.
\eeq
It too is a straightforward matter to show that this quantity follows
\beq
\frac{d {\cal Z}}{dT} = - {\rm Re}^{-1}\int |\tilde\nabla \xi|^2 dXdY.
\eeq
Thus ${\cal Z}$ always shows decay for finite Re, irrespective of the linear
shear, and, it is only
when Re = $\infty$ is ${\cal Z}$ conserved.  In the Re = $\infty$ results
we get, the enstrophy is indeed conserved between remapping events but, as
is discussed in Appendix C, the act of remapping always {\it removes} enstrophy.
The majority of this enstrophy is removed from very large wavenumber
shearwise directed Fourier modes
and since these contain very little of the energy - the
overall effect on the global energetics
are minimal.

\subsection{Results: RTG and Coherent Vortices}

\begin{figure}
\begin{center}
\leavevmode \epsfysize=10.5cm
\epsfbox{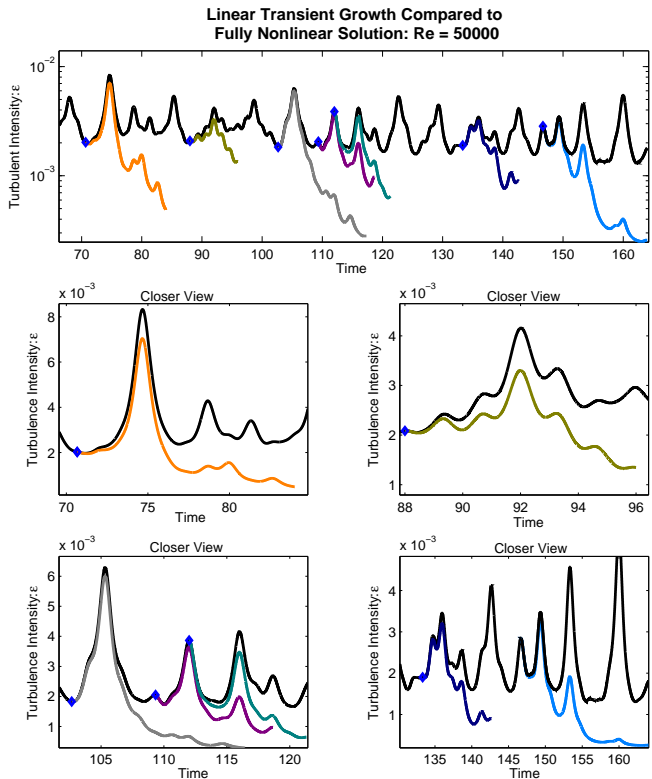}
\end{center}
\caption{{\small  Full nonlinear evolution of disturbance energy
at Re = 50000. Also plotted is the linear evolution of
disturbances whose initial condition is the fluid state of the
nonlinear solution at the specified times
 t = 70.0, 88.0, 102.6, 109.3, 112.0, 133.3, 146.7.  These linear start times are
indicated on the plots by diamonds.  Though the full solution
show strong nonlinearly,  the linear evolution
shows that the rises and falls of the total energy appear to be
dominated by the linear TG process.}}
\label{fig:TG_demo_1}
\end{figure}

We have conducted a number of runs for various values of the Reynolds number and
initial conditions (the spectral structure of the disturbance was always the same,
but we have varied the initial intensity, $\varepsilon(0)$).
Generally speaking, we have found RTG behaviour for
Reynolds numbers above some threshold. In some cases (not high enough Reynolds number
or too small disturbance strength) the RTG was transient, ultimately decaying after
a finite time (whose size depended on the two parameters mentioned above). Inviscid runs
(Re=$\infty$ nominally) seeded with initial $\varepsilon=0.01$ displayed RTG displaying
only an insignificant decay in times of the order of the run duration.
Consider first
Fig. \ref{fig:TG_demo_1}, which shows the results of a Re = 50,000 run,
demonstrating the persistence of RTG phenomenon in the time segment between $\sim 70$
and $\sim 170$.
Long after the linear evolution of
the initial conditions have died away, the nonlinear behavior shows quasi-steady
long-lived activity, characterstic of what we have called RTG. Guided by the
consequences of the Reynolds-Orr relationship (see the discussion following Eq.
\ref{reynolds-orr}) we have performed numerical experiments which demonstrate
that the peaks in the time evolution are actually driven by {\em linear} TG.
We are led to this proposition by the results of the following experiments.
At several time points during the late evolution of the flow, which are found
(in the fully nonlinear simulation) to lie just before a growing peak,
the state of the full flow is read and used as initial conditions for a
parallel numerical evaluation of the linear behavior.
This exercise reveals that the linear and nonlinear
disturbance energy fluctuations of the full system, at least locally in time (between
5 and 10 time units), are nearly the same.
Eventually the linear flow decays (as it should) while the nonlinear flow eventually
demonstrates TG again.
This quality is demonstrated in Fig. \ref{fig:TG_demo_1}
for seven separate starting points.
\par
As stated at the outset, we find that the repeated linear TG fluctuations in the
total disturbance energy
is a generic feature of the flows that have been numerically investigated
by us.  The only limitation appears to be whether or not there is
sufficient power in those set of modes of the system that are transiently growing.  For instance,
we ran a simulation (not shown) in which all the disturbance energy
was distributed amongst modes that cannot transiently grow.  The resulting nonlinear dynamics
showed fast decay exactly in line with linear theory.

It is also important to note already here that
these systems, in which sustained RTG is shown, posses an
interesting {\em spatial} characteristics. The random initial
vorticity disturbance develops into collections
of coherent long-lived vortices which, once developed, translate
in the streamwise direction with the speed of the local shear profile.
It also appears that the number of emerging structures,
showing temporal persistence, is a function of the box size in the
shearwise (i.e. disk radial) direction only.
This behavior is further detailed below.\par

\begin{figure}
\begin{center}
\leavevmode \epsfysize=3.35cm
\epsfbox{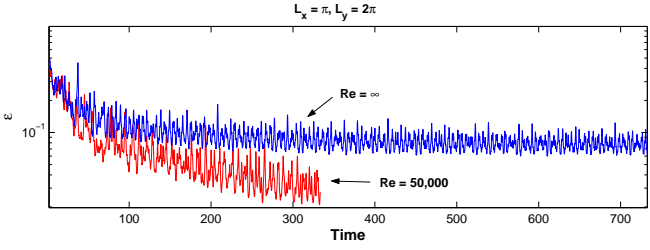}
\end{center}
\caption{{\small
The quantity $\varepsilon$ for Run a:
 white noise initial conditions for the vorticity, $L_x = 2\pi, L_y = 2\pi$.  A
 viscous (Re = 50,000) and inviscid flow, (${\rm Re} = \infty$) are depicted for comparsion.
Aside from the initial transient readjustment phase ($T < 50$),
the inviscid run demonstrates no decay over the course of its duration.}}
\label{fig:energy_Run_a}
\end{figure}

We turn now to a more detailed description of fully inviscid (save only
for numerical viscosity effect because of finite resolution and
remapping) runs. As said before, these runs show RTG persisting for the
full duration of our calculations. Three parallel evolution calculations
of three nearly identical flows at Re $= \infty$ were performed.
We refer to each of these "runs" as "a" , "b" and "c" and their parameters,
describing only differing periodic box scales, are the following:
\begin{description}
\item[\underline{Run a}:\ ] $L_x = \pi, L_y = 2\pi$,
\item[\underline{Run b}:\ ] $L_x = 2\pi, L_y = 2\pi$,
\item[\underline{Run c}:\ ] $L_x = \pi, L_y = 4\pi$.
\end{description}
Runs b and c are seeded with the initial data of Run a with some additional very weak
perturbations.
Fig. \ref{fig:energy_Run_a} depicts
the evolution of the quantity
$\varepsilon$ for Run a.
As before we see that the system demonstrates
RTG phenomenon and the energetics of the system
remain active far after the linear evolution (not shown) predicts decay.
After the initial spectrum of decaying modes has died away ($T<50$), the system's
energy reaches a quasi-steady long-term sustained level.  The same temporal features described here
are observed with Runs b and c.
\par
By comparison, we have also computed the
evolution of a viscous version of Run a with Re = 50,000.
This run demonstrated significant decay of the energy already by $T=350$ (which
was the point where the simulation was stopped).
The mix between
decay and RTG activity is clearly evident in both runs here and falls in line with implications
of the Reynolds-Orr relationship (\ref{reynolds-orr}).

\par
\begin{figure}
\begin{center}
\leavevmode \epsfysize=6.9cm
\epsfbox{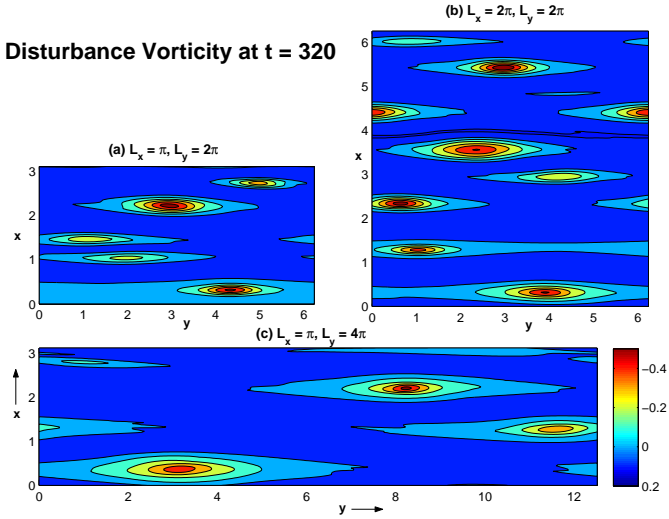}
\end{center}
\caption{{\small
The vorticity distribution at $t=320.0$
as viewed by a non-sheared observer
for Runs a,b,c at
Re = $50,000$.
In each figure a quasi steady state has been reached by the solutions. In
(a) between 3-5 vortices survive while in (b) 7-8 vortices persist.  In (c)
about 3 vortices survive.  The results are suggestive of the hyptothesis
that the number of surviving vortices is a function of the radial ($\hat x$
direction) domain size and that for any given streamline there is only one
coherent vortex associated with it.}}
\label{fig:domaincomparison_vorticity}
\end{figure}
\begin{figure}
\begin{center}
\leavevmode \epsfysize=6.9cm
\epsfbox{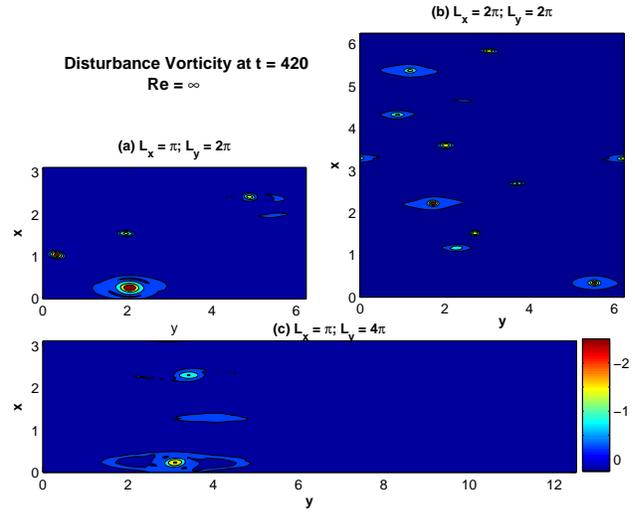}
\end{center}
\caption{{\small
Same as Fig. \ref{fig:domaincomparison_vorticity} except that
Re = $\infty$.  The concentration of vorticity is tighter than in
the viscous case.}}
\label{fig:domaincomparison_vorticity_Reinfinite}
\end{figure}

We now turn to the description of the spatial features of the results.
By $t=100$ or so, each of the three runs settle down into a quasi-steady state described as a
collection of
coherent vortical structures that propagate along with the imposed streamwise flow.
These structures
appear
(Fig. \ref{fig:domaincomparison_vorticity} \& \ref{fig:domaincomparison_vorticity_Reinfinite})
to be long-lived in that they neither transiently decay nor do they become swallowed up
by other vortices (see below) for the duration of the simulations (up to nearly $t\sim800$).
In this steady state Runs a and c support 3-5 coherent vortices while
Run b supports about 7-8 coherent vortices.
\par

Fig. \ref{fig:domaincomparison_vorticity} is representative of the long time
spatial behavior
in all the numerical simulations we have performed.  It depicts the result of
prolific vortex-vortex consumption leading to the production
a single intermediate sized coherent structure.
All simulations show that while the quasi-steady state is achieved, small to medium
sized vortices having emerged at closely neighboring values of $x$ eventually
consume each other because the background shear
invariably advects the once-separated
structures into each others vicinity.
Many structured vortices may survive into the quasi-steady phase of the evolution
but only one vortex will be associated with any streamline.
The vortex merging feature observed here is
not unexpected since this sort of thing has been
observed in other 2D numerical investigations and has been discussed at length
by McWilliams, (\cite{mcwilliams84}), Marcus (\cite{marcus93}) and more
recently by Lin {\it et al.} (\cite{lin2003}). The comparison to
these results (and thus testing our scheme)
was our motivation in running the above three runs, differing
only in the computational domains (see also below).
\par
There are two main difference in the simulations between the finite and infinite
Re cases.
First, as shown above, all finite Re simulations show
decay of their activity even
though the interim dynamics are of the RTG type seen in
the inviscid cases.
In these
latter cases (Re=$\infty$) the decay times in the total energy
are essentially
infinite since no decay was measurable {\em for the duration of the runs}.
In these cases, any decay can be caused only by the {\em numerical} dissipation attributed
to the remapping procedure (see discussion in Appendix C).
Second,
vortices appearing in the finite Re situations are
less centrally concentrated than their counterparts in the
${\rm Re} = \infty$ cases.
Fig. \ref{fig:surface_vorticity_comparisons}
shows this differing quality between two versions of Run a comparing
${\rm Re} = \infty$ and ${\rm Re} = 50,000$.
\par
\par
In summary, the results show that (a) in the comparison runs
between domains $L_y = 4\pi$ and $L_y = 2\pi$ (with $L_x = \pi$ in both),
roughly the same number of vortices survive and (b) in the comparison runs between the
domains $L_x = \pi$ and $L_x = 2\pi$ (with $L_y = 2\pi$ in both) the number of surviving vortices
is nearly double in the $L_x = 2\pi$ run.  These findings
offer qualitative support to the hypothesis, as suggested by P. Marcus
(in conversation with one of the authors), that the number of emerging coherent structures
is only a function of the size of the domain in the shearwise (disk radial)
direction and that, further,
no more than one sizeable long-lived vortex can occupy the same radial neighborhood
or zone.  Of course, the generality and veracity
of this assertion needs to be evaluated with further systematic studies.
\par


\begin{figure}
\begin{center}
\centering
\leavevmode \epsfysize=5.cm
\end{center}
\begin{center}
\leavevmode \epsfysize=6.5cm
\end{center}
\caption{{\small
Surface plot of the vorticity at long times.  Comparison between
Re = $50,000$ (top panel) and Re = $\infty$ (lower panel).
The concentration and amplitude of the vorticity in the Re = $\infty$ case is tighter than in
the viscous case.}}
\label{fig:surface_vorticity_comparisons}
\end{figure}


\section{Discussion and Summary}
\subsection{The shearing box approximation, its limits
and properties}
We have presented semi-rigorous scaling arguments in order to derive
appropriate equations
for describing small-scale dynamics in the midplane vicinity
of rotationally supported thin barotropic disks.
The procedure admits two limits, the SSB and the LSB,
whose choice depends on the length scales of interest.  In this work we
work with the SSB equations, because in this first study
we limit our interest to dynamical lengths which are small compared
to the vertical scale-height of the disk. In this way we obtain
a simpler problem, avoiding the complications resulting from
the inclusion of acoustic modes.
The SSB equations are essentially that of incompressible fluid flow
in a homogenously shearing environment with constant Coriolis parameter.  In fact,
in both the LSB and SSB scaling limits the Coriolis parameter is a local constant
and the shear is homogenous. However, by using a shearing box approximation
(which utilizes the asymptotic scaling constants $\epsilon$ and $\delta$ - see
Appendix A), a physical scale in the disk is set. It appears thus that
the argument of Balbus (2003) (regarding the above mentioned
Longaretti's speculation on the numerical resolution needed to see
instability in rotating flows) is not well founded.

The asymptotic equations describing the two shearing box scalings,
as derived in the Appendix A, relate to those used elsewhere
in the literature.  The SSB equations, in their linearized form,
are the same as those used by Chagelishvili et al.  (\cite{chag}).
Identical equations, i.e.  describing the
SSB with no a priori density fluctuations, are also equivalent to the
ones usually used in the fluid dynamics community to investigate inviscid
incompressible rotating plane Couette flow (Nagata, \cite{nagata}).
In their full nonlinear form, the equations for the SSB are equivalent to the equations
cited by Longaretti (\cite{longaretti}) where the density disturbances are everywhere zero.
\par
The equations describing the LSB admit acoustic disturbances unlike the SSB limit.
In their linearized form, the equations in the LSB are only similar to the equations
used by Tevzadze et al. (\cite{tevzadze}) where,
in their work, steady state pressure gradients are assumed to be uniform
rather than having the positional dependence they are understood to have
for barotropic pressure profiles, cf. (\ref{px}) and (\ref{pz}).
The fully nonlinear equations of the LSB are similar to those used
by BHSW except that in their treatment the steady density and
pressure profiles are considered to be uniform (avoiding
vertical structure effects). This seems to be, thus,
a kind of an intermediate limit between our SSB and LSB.
\subsection{TG: its linear decay and its nonlinear reccurrence}
\par
\par
Both the finite and infinite Re number simulations described in this work show
that long after linear theory predicts the decay of all disturbances,
the total disturbance energy exhibits repeated episodes of transient growth.
This is established by considering the
parallel linear evolution of the nonlinear state
at various points in time.  The condition of the full nonlinear flow is
used as the initial condition for the linear evolution and we
find that the linear and nonlinear development closely shadow
each other for at least 5 time units.  Where there is linear TG
the full flow also demonstrates TG.  Fig. \ref{fig:energy_Run_a}
qualitatively indicates that the temporal pattern in the RTG
is complex and appears to
be chaotic.\par

Whether a similar result can be expected in 3D, despite BHSW repeated
claims that this is impossible, must be decided by 3D simulations.
Recent simulations of 3D plane Couette flows (without
rotation) at relatively low Re-numbers by Schmiegel and Eckhardt (\cite{schmiegel})
show similar (to the one described here) behavior of the integrated domain energies.
While eventual decay is observed in most cases, the fluid response is chaotic in
the interim.   They further show that simulations with the same total initial energy
and Re numbers but with differing initial conditions
can exhibit vastly differing decay timescales - and in some instances
there is no dissipation over the duration of a simulation (Eckhardt, Marzinzik \&
Schmiegel 1998).
Though not reported here in detail, we also observe similar
sensitivity to initial conditions for simulations with finite Re numbers
in the way described in their work.  A detailed
exposition of these results will be reported in our future work.
\par

Though the detailed nonlinear feedback mechanism has not yet
been identified it seems apparent to us that the linear TG phenomenon is
a central player of this dynamical system and is a key element in the dynamics
of rotationally supported flows like thin Keplerian disks.
The findings of this work also lend support
to the conjectures put forth by Baggett, et al. (\cite{baggett}),
Waleffe (\cite{waleffe}),
 and more recently Grossmann (\cite{grossmann}):
that subcritical flows, in the sense that there
are no linearly unstable processes,  are governed by an interplay
between linear TG and some sort of nonlinear feedback of energy back onto
TG modes.

\subsection{Coherent structures}
Our results have some bearing on certain fundamental questions pertaining
to 2D turbulence.  In the work of
McWilliams (1984), in which simulations of 2D non-globally sheared
geostrophic flows were performed,
it was shown that
vorticity concentrations at intermediate scales exhibit long lifetimes: i.e.
times that are long compared to their eddy-turnaround times
and with sizes larger than the dissipation scales.
This result inspired the hypothesis that
it is a typical feature for two-dimensional
geostrophic flows to sustain these types coherent
vortices and, further, for them to retain their independence and identity
unless there is a close (and usually destructive) encounter with another vortex.     Additional
support for this view was reported
by Bracco {\it et al.} (\cite{bracco}) using very high resolution 2D simulations
of the same problem.  The appearance and persistence of vortices at intermediate
scales of the computational domain in our work qualitatively falls in line with
the suggestion that the persistence of intermediate scale vortices are a generic feature
of these sorts of flows.\par
Furthermore, the appearance of widespread merging of smaller vortices leading to
the formation of intermediate scale quasi-steady
structures is also consistent with what has been observed in the previously
mentioned non-globally sheared simulations by
McWilliams  and
Bracco {\it et al.};
and also those of the 3D shearing simulations of Lin {\it et al.} (\cite{lin2003}).
What the results of this work add to the emerging picture of such flows is that
with a globally imposed shear, intermediate sized vortices not only may appear
and persist for a long time but they also tend to be solitary.  In other words, multiple
vortices may initially emerge but
once the quasi-steady state is reached and most of the vortex-vortex merging/destruction
events have taken place,
a given point in the shearwise direction $x$ (irrespective of
any streamwise position $y$) will have only one sizable coherent vortex associated with it
.
This pattern is consistent with the speculations made to this
end by P. Marcus (private communication).
\par
We find it somewhat encouraging that the number of emerged quasi-steady vortices
increases when the domain size in the radial direction
is increased.  It suggests strongly that the emergence of coherent structures
in such flows is an intrinsic phenomenon and it is not, say for instance,
a by-product of the boundary conditions used.
By this we mean to say that if at the end of
our simulations we were to find that only one or two large scale vortices
survived into the quasi steady
state stage {\em irrespective} of the size of the computational domain in the
shearwise direction,
then we would probably feel that the pattern of developing and sustained coherence
would not be indicative of phenomena intrinsic to these flows but rather
an artifact of the boundary conditions employed
(those here being SB periodicity).
\par

\subsection{Summary}
Though the numerical results reported in this work are exclusively
of {\em two dimensional} nonlinear simulations,
we think that they may contribute to the understanding of
subcritical transition in more general shear flows as well
(including rotating ones).
We have found here support to the idea that
excepting the special circumstance, where initial conditions
are set up {\it via conspiracy} (in the sense of the phrase as used
by  Goldreich \& Lynden-Bell, \cite{goldreich65})
to have no spectral power in TG modes,
these 2D flows are likely to be dynamically active and to
exhibit and maintain robust vortices.  Indeed, in viscous 2D flows such activity
must invariably be limited to finite times, after which such activity
and the associated vortices will decay.  Nonetheless, accretion disk Reynolds
numbers are enormous - much larger than those arising from numerical roundoff in our calculations.
As such, we may expect correspondingly enormous decay times (already larger than the several hundred
box orbit times of our simulations) during which the occurrence of {\it new perturbations} can
not be excluded.

It is thus reasonable to still
seriously consider the hypothesis
that thin Keplerian disks may be hydrodynamically active,
where the activity we envisage is of the RTG kind.
It still remains to be seen, of course, whether or not
three dimensionality and its introduction of stratification and Coriolis
effects in rotating flows significantly alters the
implications of our results.

Based on the results of
their simulations, BHSW strongly advocate just that,
but our work indicates that this issue deserves additional
consideration. In particular, we are now engaged in extending
our simulations to 3D, where different numerical schemes
(spectral methods) in high spatial resolution and long
integration times are employed.
The effects of stratification and compressibility are also being addressed.
Since we have found here
that the initial conditions may be crucial in determining the
persistence time of RTG, we intend to experiment with a variety
of these and ascertain optimal perturbations.
The results of these simulations may decide the
important and still controversial issue
if complex dynamical turbulence-like activity
may be driven by purely hydrodynamical processes of the RTG
kind in rotationally supported thin circumstellar disks.

\section{Acknowledgements}
The authors thank the anonymous referee for his suggestions
that which helped us improve the presentation  of our ideas
and results.
The authors would also like to thank the Israeli Science Foundation for
making this work possible.  We also thank Jeff Cuzzi, Sandy Davis
and Denis Richard at NASA Ames Research Center
for their useful comments and for a critical suggestion they made to us.
We are grateful to Ed Spiegel and Giora Shaviv for their
willingness to discuss with us the issues of this paper and for
sharing with us their insights.

\newpage


\newpage
\onecolumn
\appendix
\section{Derivation of the Shearing Box (SB) approximation equations}

The shearing box (or sheet) (SB) approximation has been used in a variety of
astrophysical applications. It appears that the studies of Hill (1878) on
lunar dynamics have provided the basic ideas for its development.
The approximation has more recently been utilized by Goldreich \& Lynden-Bell
(1965) and Toomre (1981) in the context of galactic disks, Wisdom \& Tremaine (1988)
in the study of planetary rings, Goodman \& Ryu (1992)for accretion disks
 and by Hawley, Gammie \& Balbus (1995)
in the study of magnetized accretion disks. It is within this approximation
that most of the works mentioned in the Introduction have been done and
it is used in this and our subsequent works. It is important, in our view, that
all the approximations made in deriving the SB equations in their various
limits be precisely spelled out and clarified and this is the
purpose of this Appendix.

\subsection{Global equations in a rotating frame, cylindrical coordinates}
We start with the hydrodynamical equations, describing an ideal fluid
(actually an inviscid and isentropic flow), in cylindrical coordinates (unit vectors
$ \runit, \phunit, \zunit$)
and in a frame of reference rotating
with a given angular velocity $\Omega_0 \zunit$ (see below).
The only body
force acting on the fluid is the gravitational attraction of a point
mass $M$, located at the origin. The equation of state is assumed
to be one of a perfect gas with constant adiabatic coefficient $\gamma$.
The equations of mass momentum and energy conservation are then
\beqa
\partial_t \rho + \nabla \cdot (\rho {\bf w})&=& 0,
\label{massvecrot}
\\
\partial_t{\bf w} + ({\bf w}\cdot \nabla){\bf w}
+ 2 \Omega_0 (w_r \phunit - w_{\varphi} \runit) &=&
-\frac{1}{\rho}{\nabla P}-r \Omega_0^2\, \runit-\nabla \phi,
\label{momvecrot}
\\
\partial_t P + ({\bf w}\cdot \nabla)P + \gamma P \nabla\cdot {\bf w} &=& 0,
\label{enervecrot}
 \eeqa
where $\rho, {\bf w}$ and $P$ are the density, velocity and pressure fields
respectively and $\phi(r,z)$  is the gravitational potential which is
explicitly given by $\phi(r,z)= GM(r^2+z^2)^{-1/2}$.

Before making the crucial step in the derivation of the SB
equations (i.e. considering a small region around a point and deriving
the local equations valid approximately in this region)
we note that for any given rotation law $\Omega= \Omega(r)$, there
exists a steady axisymmetric solution, with the pressure
satisfying a global barotropic relation (i.e. being a well defined
function of $\rho$ - $P=P(\rho)$) and the velocity being composed of
the rotation only, that is ${\bf w}= r\, [\Omega(r)-\Omega_0]\,
\phunit $, as expressed in the rotating frame (see e.g. Tassoul \&
Tassoul 1978). Denoting by $\rho_b(r,z)$ the density in this solution
for a given rotation law $\Omega(r)$ we can express the gravitational
force per unit mass as follows
\beq
\nabla \phi = -\frac{1}{\rho_b} \nabla P_b + r\,
\Omega^2 \runit, \label{nablaphi}
\eeq
where $P_b(r,z)$ is obtained from $\rho_b$ by the barotropic relation. Relation
(\ref{nablaphi}) allows one to substitute for the gravity term in
(\ref{momvecrot}) and we shall do it below. It is important to notice that
if the rotation law $\Omega(r)$ is close to the Keplerian one and the disk
the fluid is cold enough (that is the vertical scale height is small),
the steady solution is an essentially rotationally supported disk.

\subsection{Local nondimensional box equations in a rotating frame, Cartesian coordinates}
We concentrate now on a small region around a point ${\bf r}_0$ -
($r=r_0$, $\varphi=\varphi_0$,
$z=0$) - , that is, some typical point in the midplane of our disk. Let the region
be of size $\Delta$, so that it (henceforth referred to as the {\em box})
is defined by
\[
r_0-\frac{\Delta}{2}\leq r \leq r_0+\frac{\Delta}{2};~~~~~
\varphi_0-\frac{\Delta}{2 r_0} \leq \varphi \leq \varphi_0 +\frac{\Delta}{2 r_0};~~~~~
-\frac{\Delta}{2}\leq z \leq \frac{\Delta}{2},
\]
where $\delta \equiv \frac{\Delta}{r_0} \ll 1$.

The smallness of $\delta$ allows the employment of Cartesian coordinates, since
locally the curvature in the azimuthal direction is only slight and, as we shall
see below, drops out at lowest order in $\delta$. Thus we can transform the coordinates
\beq
x \equiv r-r_0,~~~~~y \equiv r_0(\varphi-\varphi_0);
\label{transcord}
\eeq
and effect this transformation on equations (\ref{massvecrot}-\ref{enervecrot}),
with the gravitational term in (\ref{momvecrot}) substituted from
(\ref{nablaphi}).

Before writing out these equations in detail we nondimensionalize them,
because only then we will be able to systematically neglect terms which are of higher
order in the small parameter $\delta$. Thus $x, y, z$ are scaled by $\Delta=\delta r_0$,
$t$ by the Kepler time at $r_0$, $\tau_{K}=\Omega_K^{-1}(r_0)\equiv \sqrt{r_0^3/(GM)}$,
$\Omega$ by $\Omega_K(r_0)$, $\rho$ by $\rho_0 \equiv \rho_b(r_0,\varphi_0,0)$,
$P$ by the corresponding $P_0$ and the velocities by $v_0 \equiv
\Omega_K \Delta = \delta \Omega_K(r_0) r_0$. Note that the velocity scale
is related to the typical local sound speed scale, $c_{s0}\equiv \sqrt{P_0/\rho_0}$
by the relation
\[
v_0 = \frac{\delta}{\epsilon} c_{s0},
\]
where $\epsilon \equiv c_{s0}/[\Omega_K(r_0) r_0]$ and thus $H \equiv \epsilon r_0$
is the local vertical scale-height of the disk.
In cold disks, by definition $\epsilon \ll 1$ and thus they are thin.  We assume throughout
this discussion that disks are cold.

We do not specify for the moment the order of magnitude of the ratio of the small
parameters $\delta$ and $\epsilon$.
Their ordering will correspond, as we shall see later, to different physical regimes
of the local problem. Thus keeping the two small parameters as they are, we obtain
the following set of nondimensional equations, which are valid in the small box
around the point ${\bf r}_0$:
\beqa
\partial_t \rho + \nabla \cdot (\rho {\bf w})&=& 0,
\label{massbox}
\\
\partial_t w_x + ({\bf w} \cdot \nabla) w_x
- 2 \Omega_0 w_y - 2 q \Omega_0^2 x &=&
-\left( \frac{\epsilon}{\delta} \right)^2   \left( \frac{1}{\rho}{\partial_x P}-
\frac{1}{\rho_b}{\partial_x P_b} \right) ,
\label{momxbox}
\\
\partial_t w_y + ({\bf w}\cdot \nabla) w_y
+ 2 \Omega_0 w_x &=& -\left( \frac{\epsilon}{\delta} \right)^2
\frac{1}{\rho}{\partial_y P},
\label{momybox}
\\
\partial_t w_z + ({\bf w}\cdot \nabla) w_z &=&
-\left( \frac{\epsilon}{\delta} \right)^2  \left( \frac{1}{\rho}{\partial_z P} -
\frac{1}{\rho_b}{\partial_z P_b} \right),
\label{momzbox}
\\
\partial_t P + ({\bf w}\cdot \nabla)P + \gamma P \nabla\cdot {\bf w} &=& 0,
\label{enerbox}
 \eeqa
where we have kept only terms in lowest order in $\delta$ and
therefore neglected the curvature terms, so that the above
equations (that is, the vectors and derivatives) are actually
expressed in local Cartesian coordinates, defined in
(\ref{transcord}). The net radial body force in the rotating frame
$r(\Omega^2-\Omega_0^2)$ has been expanded around $r_0$ up to order $\delta$,
where $\Omega_0$ the rotational velocity of the frame is actually
the value of the rotational law angular velocity at $r_0$, that is,
$\Omega(r_0)$ expressed in units of the Keplerian rotational velocity
at that point. For Keplerian rotation law we obviously have $\Omega_0=1$.
The parameter $q$ is related to the Oort constant $A$:

\beq q
\equiv - \left( \frac{d \ln \Omega}{d \ln r} \right)_{r_0}=-2
\frac{A}{\Omega_0}. \label{qdef}
\eeq

In the terms on the right
hand side of the momentum equations we have the differences
between the pressure force components and the corresponding
components in the basic steady barotropic state, referred to
above.
That barotropic state is clearly axially symmetric and thus
$\partial_y P_b = \partial_y \rho_b=0$, but in the radial
and vertical directions we can not expect a priori the uniformity
the density and pressure.

Note, however, that because the ratio $(\epsilon/\delta)^2$ multiplies
the pressure force terms in
question, their size will have to be such as to allow for proper
balancing of the equations (see below).

\subsection{Removal of steady state linear shear - the shearing box (SB) equations for
the perturbations}
We notice now that ${\bf w}=2Ax {\bf \hat y}$, $P=P_b$ and $\rho=\rho_b$
is a steady solution of equations (\ref{massbox}-\ref{enerbox})
and therefore we consider this solution as our basic flow and perturb
it by adding functions of space and time (denoted by primed quantities).
Thus we write
\beqa
{\bf w} &=& 2 A x {\bf \hat y} +  {\bf u'},
\\
\rho &=& \rho_b + \rho',
\\
P &=& P_b + P',
\eeqa

Substitution of the above form of the functions in the equations gives rise to the
removal of the linear shear steady solution and the following equations for
the perturbations ${\bf u}'$, ${\rho'}$ and $P'$ are obtained:
\beqa
& &  \partial_t \rho' + 2 A x \partial_y \rho' + \rho_b \nabla \cdot {\bf u'}+
u'_z\, \partial_z\rho_b +u'_x\, \partial_x \rho_b
+ \nabla\cdot(\rho'{\bf u'}) = 0,
\label{masssbox}
\\
& & \partial_t u'_x + 2 A x \partial_y u'_x - 2 \Omega_0 u'_y +({\bf u'} \cdot \nabla)
u'_x =
-\left( \frac{\epsilon}{\delta} \right)^2
 \left( \frac{{\partial_x P'}}{\rho_b+\rho'}\,
- \frac{\rho'\partial_x P_b}{\rho_b^2+\rho_b\rho'}\,  \right),
\label{momxsbox}
\\
& & \partial_t u'_y + 2 A x \partial_y u'_y + 2 (A +  \Omega_0) u'_x +
({\bf u'} \cdot \nabla) u'_y =
-\left( \frac{\epsilon}{\delta} \right)^2
\left( \frac{{\partial_y P'}}{\rho_b+\rho'}\,\right),
\label{momysbox}
\\
& & \partial_t u'_z + 2 A x \partial_y u'_z  + ({\bf u'} \cdot \nabla) u'_z =
-\left( \frac{\epsilon}{\delta} \right)^2
 \left( \frac{{\partial_z P'}}{\rho_b+\rho'}\,
- \frac{\rho'\partial_z P_b}{\rho_b^2+\rho_b\rho'}\,  \right),
\label{momzsbox}
\\
& & \partial_t P' + 2 A x \partial_y P' + u'_x\,\partial_x P_b + u'_z\,\partial_z P_b
+ \gamma P_b \nabla \cdot {\bf u'} + {\bf u'}\cdot\nabla P'+ \gamma P' \nabla \cdot {\bf u'}= 0.
\label{enersbox}
\eeqa
where we have retained the full nonlinear terms, that is,
did not expand in the perturbations.

This set of equations for the perturbations is quite general and it includes
all the different cases which were used in the literature. The advantage of this
formulation is in its consistency and physical transparency as to which approximations
are made (see below).

\subsubsection{Small shearing box (SSB) equations ($\delta\ll\epsilon\ll 1$)}

When we chose $\delta \ll \epsilon$, the shearing box size is much smaller
that the disk's thickness (the unperturbed pressure vertical scale height).
This choice implies that the pressure force terms in equations
(\ref{momxsbox}-\ref{momzsbox}) are multiplied by a very large factor
and therefore if we want that in the prevailing dynamics these terms
be of the order of the inertial terms, they themselves must be of the
order of $(\delta/\epsilon)^2$.

Consider first the terms containing the gradient of the steady
barotropic pressure $P_b$. The $z$ derivative of $P_b$
(and thus of $\rho_b$ as well) is in
our units clearly of the order $(\delta/\epsilon)^2$, because
by nondimensionalizing equation (\ref{nablaphi}) one gets that
in the box
\beqa
\frac{1}{\rho_b}{\partial_x P_b} &=& -\left(\frac{\delta}{\epsilon^2}\right)
(1+\delta x)\left\{\Omega^2(x)- \left[(1+\delta x)^2+\delta^2
 z^2\right]^{-3/2}\right\},
\label{px}
\\
\frac{1}{\rho_b}{\partial_z P_b} &=& -\left(\frac{\delta}{\epsilon}\right)^2
z \left[(1+\delta x)^2 + \delta^2 z^2 \right]^{-3/2},
\label{pz}
\eeqa
where $\Omega(x)$ is the rotational nondimensional rotational velocity in
the box.

The order of magnitude of $\frac{1}{\rho_b}{\partial_x P_b}$ depends
on the rotation law. If the rotation law is exactly Keplerian, we have
$\Omega^2(x)=(1+\delta x)^{-3}$ and
$\frac{1}{\rho_b}{\partial_x P_b}= \order{\delta^3/\epsilon^2}$
and $\frac{1}{\rho_b}{\partial_z P_b}= -z\delta^2/\epsilon^2 + \order{\delta^3/\epsilon^2}$.
More generally
speaking, in order to have
$\frac{1}{\rho_b}{\partial_x P_b}= \order{\delta^3/\epsilon^2}$,
the basic rotation state does not have to be exactly Keplerian, but just close enough
to it.  In other words, order 1 departures from a Keplerian rotation profile
are needed to see significant radial gradients of the steady state in this approximate limit.

Thus, assuming that the rotation law is close enough to the Keplerian
(some special choices appear in the body of the paper), and for the sake
of economy of the notation, we make the replacements
\beq
\partial_i P_b \to \left( \frac{\delta}{\epsilon} \right)^2 \partial_i  P_b
\eeq
and
\beq
\partial_i \rho_b \to \left( \frac{\delta}{\epsilon} \right)^2 \partial_i \rho_b
\eeq
for $i=x,z$, remembering that the units of the barotropic profile spatial derivatives
have been scaled accordingly.

Examine next the terms containing the perturbations $P'$ in
equations (\ref{momxsbox}-\ref{momzsbox}). If these are not
sufficiently small (that is of order $\delta^2/\epsilon^2$) we may formally
write
\beq
P'= P'_0 + \left(\frac{\delta}{\epsilon}\right)^2 p +...
\eeq
With this expansion (and using the above scaling of the barotropic terms),
equations (\ref{momxsbox}-\ref{momzsbox}) give in lowest order
\beq
\nabla P_0'=0
\eeq
that the $\order{1}$ part of the pressure perturbation is uniform in the box.
Using this in equation (\ref{enersbox}) gives in lowest order
\beq
\partial_t P'_0 = -\gamma \nabla\cdot[(P_b+P'_0){\bf u'}].
\eeq
Integration of this equation over the volume of the box shows that the
time variation of $P'_0$ depends on the boundary conditions. For both periodic
or zero velocity boundary conditions (which are usually employed) we get
that $P'_0$ is also constant in time. Thus this constant pressure shift
may be formally absorbed into $P_b$ and consequently changing nothing.  Thus we may now
formally write
\beq
P' \equiv \left( \frac{\delta}{\epsilon} \right)^2 p,
\label{ppert}
\eeq
in which $p$ is now understood to be an order one quantity. This actually
amounts to redefining the units of the pressure perturbation.

Substitution of these scalings into (\ref{masssbox}-\ref{enersbox}) and dropping
terms of $\order{(\delta/\epsilon)^2}$ and smaller gives that
the energy equation (\ref{enersbox}) simplifies to
\begin{mathletters}
\beq
 \nabla \cdot {\bf u'} = 0,
\label{enerssbox}
\eeq
that is, we are dealing in this approximation with an incompressible flow.
This is physically consistent with the fact that the relative pressure
perturbation was assumed to be very small. We have an incompressible flow
with the density of any fluid element conserved by the flow and the
the small pressure perturbations describe the dynamics of non-acoustic
(that is solenoidal, or vortical) modes only.

The three momentum equations become:
\beqa
\partial_t u'_x + 2 A x \partial_y u'_x - 2 \Omega_0 u'_y + ({\bf u'} \cdot \nabla)
u'_x &=& -\left( \frac{1}{\rho_b+\rho'}\,{\partial_x p}
- \frac{\rho'}{\rho_b^2+\rho_b\rho'}\, \partial_x P_b \right)
\label{momxssbox}
\\
\partial_t u'_y + 2 A x \partial_y u'_y + 2 (\Omega_0+A) u'_x + ({\bf u'} \cdot \nabla)
u'_y &=& -\left( \frac{1}{\rho_b+\rho'}\,{\partial_y p}\right),
\label{momyssbox}
\\
\partial_t u'_z + 2 A x \partial_y u'_z  +({\bf u'} \cdot \nabla) u'_z &=&
-\left( \frac{1}{\rho_b+\rho'}\,{\partial_z p}
- \frac{\rho'}{\rho_b^2+\rho_b\rho'}\, \partial_z P_b \right)
\label{momzssbox}
\eeqa
Finally, the equation for the density perturbation $\rho'$
results from (\ref{masssbox}) and, in this order, reads
\beq
\partial_t \rho'  + 2 A x \partial_y \rho' + {\bf u'}\cdot \nabla \rho'= 0.
\label{massssbox}
\eeq
\end{mathletters}
The set (\ref{enerssbox}-\ref{massssbox}), supplemented by the appropriate
boundary conditions (usually chosen to be periodic), describes the
full nonlinear problem in the SSB approximation, that is, for $\delta\ll \epsilon$.
These equations with $\rho' = 0$ throughout is equivalent to the system
studied by Longaretti 2002 and is often times
referred to and studied as rotating plane Couette flow (Nagata, \cite{nagata}).
The linear problem contains the one studied
by Chagelishvili {\em et} al. (2002).

\subsubsection{Large shearing box (LSB) equations ($\delta \sim \epsilon \ll 1$)}

We chose now $\delta = \epsilon$ the size of the shearing box is of the order of the
vertical thickness of the disk. In this case
the relative size of pressure perturbations is of the order of
the relative density perturbations (to balance the relevant terms) and
the vertical and horizontal derivatives of the barotropic pressure
are necessarily retained in all of the equations as well.
When the flow is nearly Keplerian, only the vertical
derivatives make it in at the lowest order as it was argued
in the previous section.
We get equations identical to the full set
(\ref{masssbox}-\ref{enersbox}) but with
$\delta=\epsilon$ and where, in practice, only the first order terms of the basic state
barotropic profiles (\ref{px}-\ref{pz}) survive to this lowest order.
Unlike the SSB, in this limit
acoustic waves will be included in the equations as the flow is not incompressible.
We shall treat this system in future works.\par
It should be noted that
Tevzadze {\em et.} al (2003) used a simplified version of the linear limit of these equations.
The system studied by Balbus, Hawley and collaborators (for instance
Eqs. 3.5a-b in Balbus, Hawley and Stone, \cite{bhs96}) is a special case
of this problem with the background barotropic density and pressure
profiles assumed to be uniform.

\section{Details of the nonlinear 2D numerical method}\label{numerical_methods}
We begin with
(\ref{fullxieqn}) and (\ref{streamvortyrel}).
 We assume all quantities to be Fourier-Galerkin expanded in both the $X$
 and $Y$ directions. For example we have for
 the vorticity,
 \beq
 \xi = \sum_{k,\ell}\xi_{_{k\ell}}(T)e^{ikX + i\ell Y};
\label{fourier_galerkin_expansion}
 \eeq
we similarly expand in this way for the stream function $\psi$.
Note that each
In the Fourier basis the equations
are
\begin{mathletters}
 in which the Jacobian $J$ is
 \beq
 J(\psi,\xi) = \partial_{_Y}\psi(\partial_{_X}\xi + q T \partial_{_Y}\xi)
 - \partial\xi_{_Y}(\partial_{_X}\psi + q T \partial_{_Y}\psi).
 \eeq
 \end{mathletters}
 Like $\xi_{_{k\ell}}$, $J_{_{k\ell}}$ references the $k,\ell$ Fourier
 component of the Jacobian.
The time stepping is done via a modified Crank-Nicholson method
outlined below.  We write the above equation
in the more generalized form,
\beq \frac{d \xi_{_{k\ell}}}{d T} = -J_{_{k\ell}} + {\cal L}_{_{k\ell}}(T)\xi_{_{k\ell}},
\label{evolution_eqn} \eeq
where the linear operator
${\cal L}_{_{k\ell}}$ acting on Fourier component $\bf k$ has
an explicit $T$ dependence.  The evolution scheme time centers
the evaluation of the nonlinear term $J_{_{k\ell}}$ and uses a time averaged
evaluation of the linear operator.  If superscript $n$ refers to
the $n^{{\rm th}}$ time step and if $\delta t$ refers to the time
difference between successive time steps then the temporal
discretization of (\ref{evolution_eqn}) becomes, \beq
\frac{\xi^{n+1}_{_{k\ell}} - \xi^{n-1}_{_{k\ell}}}{2\delta t} =
\frac{{\cal L}_{{\bf k\ell}}(T+\delta t)\xi^{n+1}_{_{k\ell}} +{\cal L}_{{\bf k\ell}}(T-\delta
t)\xi^{n-1}_{_{k\ell}} }{2} - J^{n}_{_{k\ell}}(T),
\eeq
reshuffling
terms gives,
\beq \xi^{n+1}_{_{k\ell}} = \frac{1 + \delta t {\cal
L}_{_{k\ell}}(T-\delta t)} {1 - \delta t {\cal L}_{_{k\ell}}(T+\delta
t)}\ \xi^{n-1}_{_{k\ell}} - \frac{2\delta t}{1 - \delta t {\cal
L}_{_{k\ell}}(T+\delta t)} J^{n}_{_{k\ell}}(T).
\label{discritization_step_1} \eeq
We also remind the reader that $J^{n}_{_{k\ell}}$ is the Jacobian
evaluated at the time step $n$. We notice that the
terms involving the linear operator appearing in the above expressions
are the first terms of a Taylor series expansion of their
exponential forms.  That is,
\beq
\exp\left\{ \int_T^{T+\delta t} {\cal L}_{_{k\ell}} dT'\right \}
\approx 1 + \delta t {\cal L}_{_{k\ell}}(T)+ \order{\delta t^2},
\label{pade_appx} \eeq
we replace all appearances of ${\cal L}_{_{k\ell}}$  with the above approximate form.
Thus (\ref{discritization_step_1}) becomes
\beq \xi^{n+1}_{_{k\ell}} = \exp\left\{ \int_{T-\delta t}^{T+\delta
t} {\cal L}_{_{k\ell}} dT'\right \} \xi^{n-1}_{_{k\ell}} - 2\delta t
\exp\left\{ \int_{T}^{T+\delta t} {\cal L}_{_{k\ell}} dT'\right \}
J^{n}_{_{k\ell}}(T). \eeq
Replacing the linear operators as we have
here is akin to a P\`ade Approximation (Bender \& Orszag, \cite{bender}) except in reverse: we have
``inferred'' a more accurate form of the operator by recognizing
that the fractional terms in (\ref{discritization_step_1}) are the
approximations of the exponentials which are the exact solutions
to problems with such operators.  This procedure more accurately
represents the linear evolution and it removes the well-known
high-wavenumber inaccuracies (though stable) associated with
Crank-Nicholson methods: in particular, as originating from the
denominators of (\ref{discritization_step_1}).
\section{Remapping: its rationale and technical details}
The original call to our attention of the ideas behind {\em
remapping} is credited to P.S. Marcus who, in several private
comminications with one of the authors (OMU), sketched out the
essential rudiments of the problems (and their resolution) as they
are discussed herein.  To our knowledge, Cabot (\cite{cabot}) is the only published work
in the literature in which
the steps describing the remapping procedure is both outlined and discussed with some detail.
\subsection{The rationale}

We have two goals in mind when developing
(numerical) nonlinear solutions of (\ref{fullxieqn}) and
(\ref{streamvortyrel}):
\begin{enumerate}
\item to accurately implement periodicity in the SC frame (as in the programme outlined
in Hawley et al.,
\cite{hgb}) and,
\item to resolve spatially coherent structures for as long a time as possible.
\end{enumerate}
The first goal is automatically achieved by going into the SC
frame and by assuming a spatially doubly periodic Fourier-Galerkin
decomposition of the solution (see Appendix B).\par The trouble
lies in the inability of resolving spatial coherence for an arbitrarily long time in the SC
frame.
To appreciate why this a problem in the first place,
we must first define what we mean when we speak of spatial coherence. For a
non-sheared normal observer, i.e. someone in the NSC frame, we understand
a {\it coherent structure} to be a fluid structure that has a
more-or-less definite size and shape.  By {\it long-lived} we
imply that the coherence in the structure remains constituted
as such for a long time, at least in the statistical sense.
In the language of this article, it is to say that in the NSC
frame the coherent structure will have a statistically-steady
wavenumber bounded spatial spectral profile with power peaking at some
finite set of Fourier wavemodes.  For the purpose of this
argument, the mechanism behind the statistically steady maintenence
of power at these wavemodes is inessential.
For the sake
of transparency in the subsequent discussion, let us suppose that all
the power resides exactly with one wavemode with wavenumber ${\bf
k_0} = ( k_0, \ell_0)$ in the NSC frame.
\par
The problem centers around the temporally limited ability to represent
in the SC frame an object that appears (statistically) steady in the NSC frame.  As time progresses
the representation of a
coherent structure in the SC frame steadily degrades fundamentally because the computational
resolution in the SC frame is limited.  Below we describe in more detail the origins
and nature of this problem.\par
To express these ideas in more mathematical terms we begin with a coherent structure whose
vortical power is concentrated, as we said,
in a single Fourier wavemode with
wavenumber $k_0,\ell_0$ expressed in the NSC frame as,
\beq
\xi(x,y,t) \sim \xi_{_{k_0 \ell_0}}e^{i k_0 x + i \ell_0 y}, \label{vorty_nsc}
\eeq
in which $\xi(x,y,t)$ is the representation of the vorticity in the NSC frame and
where $\xi_{_{k_0 \ell_0}}$ is the complex amplitude of the vorticity at wavemode $k_0,\ell_0$.
Because the dynamics are steady we assume that $\xi_{{k_0\ell_0}}$ is time-constant
\footnote{Even this assumption is too restrictive.  The amplitude could have some time dependence in the
argument that follows.  For the argument presented, it is enough that the power remains contained in this single wavemode,
or at worst, a handful of wavenumber bounded wavemodes.}.
Let ${\bf \Pi^{-1}}(t_{_0})\ \otimes$ symbolically represent the act of transforming from
the NSC to the SC frame  at
the reference time $t_{_0}$
(i.e. the mapping defined by the inverse of Eq. \ref{shearedmetrictransform}) and let
${\bf \Pi}(t_{_0})\ \otimes$ represent the forward transformation.
Letting $\xi(X,Y,T)$ represent the vorticity in the SC frame,
then the forward and backward coordinate mapping procedure on the vorticity
is formally the following,
\[
{\bf \Pi^{-1}}(t_{_0})\otimes\xi(x,y,t) \longrightarrow \xi(X,Y,T), \qquad
{\bf \Pi}(t_{_0})\otimes\xi(X,Y,T)
\longrightarrow \xi(x,y,t).
\]
Specifically speaking, applying ${\bf \Pi^{-1}}(t_{_0})$ on the vortical mode in (\ref{vorty_nsc})
at some starting reference time $t_{_0}$ in the NSC frame gives,
\beq
\xi(X,Y,T) = \xi_{_{k_0 \ell_0}}
 e^{i (k_0-qT\ell_0) X + i \ell_0 Y} = \xi_{_{k_0 \ell_0}}
 e^{ik_{{\rm eff}} X + i \ell_0 Y},
\qquad k_{{\rm eff}} \equiv k_0-qT\ell_0,
\label{vorty_sc}
\eeq
where $k_{{\rm eff}}$ is an effective $X$ direction wavenumber whose magnitude eventually grows with
 $T$.  Since $t_{_0}$ is the reference time for the transformation,
$k_{{\rm eff}}$
and $k_0$ are initially identical because $T=0$ when the result of the
transformation is viewed at $t=t_{_0}$.\par
It is easy to see that the coherent structure, as viewed from the SC frame,
needs to be represented by an increasingly
larger set of Fourier modes in the $X$ direction.
But because we are computationally limited to finite resolution (even in the SC frame)
eventually one will run out Fourier modes in the SC frame that can properly represent the
coherent structure.
We assume that in the SC frame the $X$ domain is represented by
a finite set of wavenumbers, $k_i$, lying within and bounded by a maximum
wavenumber $\pm k_N$, or, in other words, $-k_N < k_i < k_N,$ with $k_N > 0$.
The bounding wavenumber, $k_N$,
is set by the number of
grid points (say $N$) in the $X$ direction of the domain.
\footnote{Naively speaking: if the size of the domain in the
$\hat X$-direction is $L_x$ then roughly speaking $k_N = N/2L_x$.}
The magnitude of $k_{{\rm eff}}$ will exceed $k_N$
after a time $T_{{\rm res}}$
given by
\[
T_{{\rm res}} =  \frac{k_0}{q\ell_0} +  \frac{k_N}{q|\ell_0|}.
\]
For $T>T_{{\rm res}}$ the numerical simulation in the
SC frame requires there to be a Fourier mode $|k| = |k_{{\rm eff}}| > k_N$
in order to represent the coherent structure.  Because this mode is not there the
coherent structure can no longer be represented within the physical construction
of the computation.
We refer to this effect as
permanent {\it information-loss} which is, in short, very bad.
By contrast, with infinite spatial resolution
at our disposal (i.e. $k_N = \infty$) this information-loss phenomenon
would not occur because there would always be
a $k = k_{{\rm eff}}$ represented in the SC frame for all $T>0$.
\par
In general a coherent structure will be represented by a number of Fourier modes
and not the single one assumed for illustration here.  Nevertheless the effects
just described applies to them as well since the representation of their
power in the SC frame will drift in wavenumber according to the process just described.  Thus,
in considering a resolution of this dilemma, we must always keep in mind that the
coherent structure(s) of interest usually have some compact spectral representation
in Fourier space.
In desiring to avoid major information loss about the coherent structure(s), we mean
to protect and preserve as much
as possible the constituents of its power spectrum while it
is represented and dynamically evolved in the SC frame.\par
One final observation is required before the solution may be apparent.
Irrespective of resolution matters, one may also,
after any passage of time $\Delta t$,
apply the coordinate transformation ${\bf \Pi}(t_{_0})$
back upon the solution evolved in the SC frame and recover the original steady
coherent structure with wavemode $(k_0,\ell_0)$
in the NSC frame.  In other words the transformation,
\[
{\bf \Pi}(t_{_0})\otimes\xi(X,Y,\Delta t) \rightarrow \xi(x,y,t_{_0}+\Delta t),
\]
recovers the original coherent structure in the NSC frame.  For our particular example
we would have
\beq
\xi(x,y,t_{_0}+\Delta t) = \xi(x,y,\tilde t_{_0}) =
\xi_{_{k_0 \ell_0}}e^{i k_0 x + i \ell_0 y},
\qquad
\tilde t_{_0} \equiv t_{_0}+\Delta t.
\label{vorty_nsc_later}
\eeq
One may then restart the evolution in the SC frame by applying ${\bf \Pi^{-1}}(\tilde t_{_0})$
upon $\xi(x,y,\tilde t_{_0})$ {\em noting} that $\tilde t_{_0}$ is the new reference time,
which is $T_{{\rm rm}}$ after the original reference time $t_{_0}$.
This means that the simulation in the SC frame begins anew with {\em $T$ again set to zero}
and it is valid again for $T \ge 0$.  We call this {\em remapping}.
In terms of the example
we are using for illustration, the single wavemode vorticity again starts off with
its power in the wavemode $k_{{\rm eff}} = k_0$ as before.
The benefit of repeated applications of this procedure (of evolving and remapping)
is that coherent structure(s) can be indefinetly represented in the SC frame so long
as the remapping step is done well before the resolution loss time
is approached in the SC frame.
In our example it means that the single wavemode vorticity can be represented
for as long as one wishes as long as $T_{{\rm rm}} < T_{{\rm res}}$.
Without the act of repeated remappings the coherent structure can only be
represented up to the point $T=T_{{\rm res}}$ and no longer.
\footnote{
Mathematically speaking, the basis for why
this is all acceptable lies in the fact that
the transformation $t \rightarrow t +
\Delta t$, where $\Delta t$ is a constant, leave invariant the fundamental
equations (\ref{fullxieqnnsc}-\ref
{streamvortyrelnsc}).}
\par
The way out of this information-loss dilemma makes use of the observations in the
previous paragraph.
Starting from some time $t_{_0}$,
one may perform the numerical computation in the SC frame up until a time $T_{{\rm rm}}$,
which is much less than the time $T_{{\rm res}}$ when significant
information-loss sets in.  At $T_{{\rm rm}}$ the solution generated in the SC is
coordinate mapped into the NSC using ${\bf \Pi}(t_{_0})$
(i.e. Eq. \ref{shearedmetrictransform}).
Once back in the NSC we immediately reapply the inverse transformation
${\bf \Pi^{-1}}(t_{_0}+T_{{\rm rm}})$
(i.e. the inverse of Eq. \ref{shearedmetrictransform} with the
reference time set to $t_{_0}+T_{{\rm rm}}$) upon the solution in
the NSC frame to
begin the simulation again in the SC frame.  The evolution of the solution
in the SC frame proceeds with $T$ reset to zero again.
If there are intrinsic steady coherent structures
in the flow, this act of  repeated remappings always corrals their wave power
into a wavenumber range that  {\em is always resolved in the SC frame}; thereby
preventing the information-loss
just outlined.  It is very important, then, to apply this remapping procedure
well before any significant information about the coherent structure is permanently lost,
i.e. make sure to ensure that $T_{\rm rm} \ll T_{\rm res}$.
\par
We summarize the procedure below:
\begin{description}
\item[(i)] Vorticity data in the NSC frame at time $t_{_0}$, i.e. $\xi(x,y,t_{_0})$,
is transformed
into the SC frame via the coordinate map:
\[{\bf \Pi^{-1}}(t_s) \otimes \xi(x,y,t_{_0})
\longrightarrow \xi(X,Y,0). \]
\item[(ii)] The vorticity in the SC frame is evolved from $T=0$ to $T=T_{{\rm rm}}$.
\item[(iii)]  The solution of the vorticity at time $T=T_{{\rm rm}}$ in the SC frame
is expressed in terms of what it would look like in the NSC frame:
\[{\bf \Pi}(t_s) \otimes \xi(X,Y,T_{{\rm rm}}) \longrightarrow
\xi(x,y,t_{_0}+T_{{\rm rm}}).
\]
\item [(iv)] The reference time is updated from $t_{_0} \rightarrow t_{_0} + T_{{\rm rm}}$
and the steps repeat again with (i).
\end{description}
The choice for $T_{{\rm rm}}$ is technically arbitrary.  However, the natural choice is to
use those moments in time when the solution appears exactly periodic in the (shearwise)
$x$ direction in the NSC frame.  This is detailed in points 1-4 in section 4.1 of the text.\par
\subsection{Some technical details}
The solution for $\xi$ in the SC frame is represented as in (\ref{fourier_galerkin_expansion}).
The wavenumbers in the $X$ and $Y$ direction respectively are taken to be
\beq
k = \frac{2\pi n_{_k}}{L_x}, \ \ n_{_k} \in 0,\pm 1,\pm 2,\cdots,\pm n_{max}, \qquad
\ell = \frac{2\pi m_{_\ell}}{L_y}, \ \ m_{_\ell} \in 0,1,2,\cdots m_{max}.
\eeq
Because of the complex conjugate symmetry of the solutions in wavenumber space
we need only consider half the modes (the positive ones including zero) in the $Y$ direction.
Because this is a numerical simulation we are limited to a maximum number of modes to
be $n_{max}, m_{max}$ in the $X$ and $Y$ directions respectively.
Between remappings the maximum dynamically resolved wavenumbers are $\pm k_{max}, \ell_{max}$
given by,
\[
k_{max} \equiv \frac{2\pi n_{max}}{L_x},\quad
\ell_{max} \equiv \frac{2\pi m_{max}}{L_y},
\]
and where, of course, $k_{max} > 0$.
The vorticity is evolved from $t_{_0}$ to time $t_{_0} + T_{rm}$ whereupon we have
\beq
\xi(t_{_0} +T_{rm}) = \sum_{k,\ell}\xi_{{_{k\ell}}}(t_{_0} + T_{rm}) e^{ikX + i\ell Y}.
\label{SC_solution_at_Trm}
\eeq
At this stage the solution is considered to be periodic in the non-shearing frame.  Thus we reexpress
(\ref{SC_solution_at_Trm}) in the non-shearing frame by using the transformation
according to (\ref{shearedmetrictransform}) to find,
\beqa
\xi(t_{_0} + T_{rm}) &=& \sum_{k,\ell} \xi_{_{k\ell}}(t_{_0} + T_{rm})e^{ikx + i\ell y + iqT_{rm} \ell x } \nonumber \\
&=&
\sum_{k,\ell} \xi_{_{k\ell}}(t_{_0} + T_{rm})e^{i(k+\Delta k)x + i\ell y},
\label{solution_reexpressed}
\eeqa
in which
\beq
\Delta k = m_{_\ell} \frac{2\pi}{L_x}.
\label{kwave_shift}
\eeq
Now because at the remap time the two coordinate systems are coincident
we may reexpress the solution (\ref{solution_reexpressed}) in
the SC frame,
\beq
\xi(t_{_0} + T_{rm}) = \sum_{k,\ell} \xi_{_{k\ell}}e^{i(k+\Delta k)X + i\ell Y},
\label{remapped_SC}
\eeq
with the reference time $t_{_0}$ (cf. \ref{shearedmetrictransform})
updated by a factor $T_{rm}$: in other words,
at this point $t_{_0} \rightarrow t_{_0} + T_{rm}$.
A momentary inspection of (\ref{remapped_SC}) reveals that
at the time of remapping, in which the SC and NSC frames are coincident,
wholesale shift of power occurs {\em from} leading, transiently growing
modes {\em towards} trailing, decaying modes.  \footnote{Recall from linear theory that
for wavemodes with $\ell > 0$, transient growth is expected for modes with $k<0$ and decay
for $k>0$.}
For example supposing our
computational domain was such that $L_x = L_y = 2\pi$, if we had
a wavemode with $k=-2$ and $\ell = 3$ then, according to (\ref{remapped_SC}),
after the remap that wavemode
would become $k=1$ and $\ell =3$ based on the shift required by (\ref{kwave_shift}).
As we see,  in the updated SC basis,
subseqent each remapping  event
the reference time $t_0$ (cf. \ref{shearedmetrictransform}) accrues an additional
factor of $T_{rm}$.  Most critically,
if the remap shifts power to be outside the trailing edge maximum
wavenumber, $+k_{max}$, that power is considered
lost to the system.
\par
The following illustrations, beginning with the example in the previous
paragraph, should help clarify all of this.
Suppose that $k_{max} = 10$ and we are considering power in horizontal wavemodes
with $\ell = 3$.  As in the previous example, power in
wavemode $\xi_{_{-2,3}}$ before remapping becomes $\xi_{_{1,3}}$ after
the remap according to (\ref{kwave_shift}-\ref{remapped_SC}).
Similarly $\xi_{_{5,3}} \rightarrow \xi_{_{8,3}}$.  Near the trailing edge limit,
for example, $\xi_{_{9,3}}$ would become $\xi_{_{12,3}}$, but because the maximum positive
(trailing edge) $k$ wavenumber
is $10$ the power in the mode $\xi_{_{9,3}}$ gets shifted out of the computation all together
and it is, in this sense, totally lost.  On the other hand suppose we were considering
the wavemode $\xi_{_{-10,3}}$: after remapping its power shifts into wavemode $\xi_{-7,3}$.  After
remapping, however, the $\xi_{_{-10,3}}$ wavemode is initiated with zero power.  In fact, for the $\ell = 3$
wavemodes, the modes $\xi_{_{-10,3}},\xi_{_{-9,3}},\xi_{_{-8,3}}$ are all initiated with zero power.
Had we been considering instead wavemodes with $\ell = 5$ then
after the remapping the wavemodes
$\xi_{_{-10,5}},\xi_{_{-9,5}},\xi_{_{-8,5}},\xi_{_{-7,5}},\xi_{_{-6,5}}$ would all be set to zero.
\par
It should be clearly evident from this that a large amount of enstrophy will be lost
to the system after each remapping event.  Because we can predict exactly which wavemodes
prior to remapping will have its power shifted out of the computation, we can (and do) predict
exactly how much enstrophy (and, for that matter, energy) will be lost after each remap.  In fact, the numerical method employed
in evolving the equations here is such that between remapping events the enstrophy is exactly
conserved.  However,
what is more important
is that after a remap power is explicitly shifted towards
more positive values of $k$.
Additionally after a remap, the most extremely transiently growing modes (modes
with very large and negative $k$ wavenumbers) become initiated with zero power after remapping.
Because this remapping procedure, as prescribed here, is executed
in the Fourier-Galerkin basis and not
in physical space, {\em there is no potential for aliasing errors here}.
In other words, the remapping will not artificially create power in transiently growing modes.
This means that we can be assured that any power that develops
in transiently growing modes do so naturally (i.e. are truly physical)
via the usual three wave
interactions understood for such systems like this 2D inviscid incompressible flow.
\end{document}